\documentclass[onecolumn,authoryear]{els-mrw} 

\usepackage{amsmath,amssymb,amsfonts,amsthm,makeidx,graphicx}
\usepackage{txfonts}
\usepackage{helvet}
\begin{document}

\chapter{Habitability and Biosignatures}\label{chap1}

\author[1]{Nikku Madhusudhan}%

\address[1]{\orgname{University of Cambridge}, \orgdiv{Institute of Astronomy}, \orgaddress{Madingley Road, Cambridge CB3 0HA, UK}}

\long\def\symbolfootnote[#1]#2{\begingroup
\def\thefootnote{\fnsymbol{footnote}}\footnote[#1]{#2}\endgroup} 
\def\aj{AJ}
\def\araa{ARA\&A}
\def\apj{ApJ}
\def\apjl{ApJ}
\def\apjs{ApJS}
\def\apss{Ap\&SS}
\def\aap{A\&A}
\def\aapr{A\&A~Rev.}
\def\aaps{A\&AS}
\def\aplett{Astrophys.~Lett.}
\def\icarus{Icarus}
\def\jcp{J. Chem. Phys.}
\def\jgr{J. Geo. Res.}
\def\mnras{MNRAS}
\def\na{New A}
\def\nar{New A Rev.}
\def\nat{Nature}
\def\pasj{PASJ}
\def\pasp{PASP}
\def\planss{P\&SS}
\def\ssr{Space~Sci.~Rev.}

\maketitle

\begin{abstract}[Abstract]
The search for life beyond the solar system is a central goal in exoplanetary science. Exoplanet surveys are increasingly detecting potentially habitable exoplanets and large telescopes in space and on ground are aiming to detect possible biosignatures in their atmospheres. At the same time, theoretical studies are expanding the range of habitable environments beyond the conventional focus on Earth-like rocky planets and biosignatures beyond the dominant biogenic gases in the Earth's atmosphere. The present work provides an introductory compendium of key aspects of habitability and biosignatures of importance to the search for life in exoplanetary environments. Basic concepts of planetary habitability are introduced along with essential requirements for life as we know it and the various factors that affect habitability. These include the requirement for liquid water, energy sources, bioessential elements, and geophysical environmental conditions conducive for life. The factors affecting habitability include both astrophysical conditions, such as those due to the host star, as well as planetary processes, such as atmospheric escape, magnetic interactions, and geological activity. A survey of different types of habitable environments possible in exoplanetary systems is presented. The notion of a biosignature is presented along with examples of biosignatures on Earth and their applicability to habitable environments in exoplanetary systems. The desired properties of an ideal biosignature are discussed, along with considerations of the environmental context and chemical disequilibria in the assessment of biosignatures in diverse environments. A discussion of current state-of-the-art and future prospects in the search for habitable conditions and biosignatures on exoplanets is presented. 
\end{abstract}

\begin{keywords}
Exoplanets, habitability, biosignatures, atmospheres, spectroscopy 
\end{keywords}

\begin{glossary}[Glossary]
\term{Autotrophs}: Organisms that produce their own food using  inorganic sources.

\term{Bond albedo}: Fraction of incident stellar radiation that is reflected by a planet. 

\term{Atmospheric escape:} The escape of atoms or ions from the upper atmosphere of a planet to space.

\term{Biosignature:} Any observable aspect of a habitable environment, involving single or multiple features, which indicates the presence of life. It is used interchangeably with the term ``biomarker" which typically refers to a single molecule that is indicative of a biological process.

\term{Biogenic:} Produced through a biological process.

\term{Chemical disequilibrium:} The state of a system, such as a planetary atmosphere, in which concentrations of certain chemical compounds are not in thermochemical equilibrium for the corresponding temperature and pressure.

\term{Chemotrophs:} Organisms that derive their energy through the oxidation of chemical compounds found in their environment.

\term{Continuous habitable zone :} Region around a star where an Earth-like planet would be able to sustain liquid water on some part of the surface over geological timescales e.g. several Gyr.

\term{Extremophile:} An organism that can live in extreme  environments.

\term{Exoplanet:} A planet located outside the solar system.

\term{Geochemical cycling:} The cyclic transfer of chemical elements through an environment, which can involve interactions between the atmospheric, geological and biological processes. 

\term{Effective temperature:} The temperature corresponding to the total radiation emitted by a body assuming the spectral energy distribution to be the Planck function, i.e., that of a blackbody. 

\term{Habitability:} The condition of an environment which satisfies the essential requirements for sustaining life. 

\term{Habitable zone:} The region around a star where a planet with certain atmospheric conditions can sustain liquid water on the surface. 

\term{Heterotrophs}: Organisms that rely on other organisms for nutrition.

\term{Hycean worlds:} Temperate planets with ocean covered surfaces and H$_2$-rich atmospheres that can sustain habitable conditions similar to those in Earth's oceans.

\term{Methanogens:} Organisms that produce methane as a result of their metabolism. The process through which these microorganisms generate methane is called methanogenesis.

\term{Orbital separation:} The distance between a planet and its host star, typically measured in astronomical units (au).

\term {Outgassing:} The release of gases from a planet's interior to its surface and atmosphere, usually by volcanism.

\term{Phototrophs:} Organisms that derive their energy from light.

\term{Stellar luminosity:} The total amount of energy emitted by a star per unit time. 

%\term{Stellar radiation:} Energy emitted by a star in the form of electromagnetic radiation, i.e. light. 

\term {Sub-Neptunes:} Exoplanets with sizes smaller than Neptune with radii typically between 1.5-4 R$_ \oplus$. 

\term {Transiting exoplanet:} An exoplanet that transits in front of its host star periodically as seen by a telescope. 

\term{Transmission spectrum:} The transit depth of a transiting planet as a function of wavelength. 

\end{glossary}

\begin{glossary}[Nomenclature]
\begin{tabular}{@{}lp{34pc}@{}}

ATP & Adenosine triphosphate \\
au & astronomical unit \\
CHNOPS & Carbon, hydrogen, nitrogen, oxygen, phosphorus,
and sulphur\\
CHZ & Continuous habitable zone \\
DMS & Dimethyl sulfide \\
HZ & Habitable zone \\
JWST & James Webb Space Telescope \\
ppm& Parts per million by volume\\
ppb& Parts per billion by volume\\
SED& Spectral Energy Distribution %\\

\end{tabular}
\end{glossary}

\begin{glossary}[Learning objectives]
This chapter is aimed at advanced undergraduate students and early career researchers with a basic background in natural sciences. The chapter provides an introductory compendium of key aspects related to planetary habitability and biosignatures as applicable to extrasolar planets. The topics covered include planetary habitability, factors affecting habitability, examples of potentially habitable environments beyond Earth, biosignatures, examples of terrestrial biosignatures, current status and future prospects for biosignature detection on exoplanets. Key concepts and examples are introduced along with references to literature for more detail. 

\end{glossary}

\section{Introduction} \label{sec:intro}
\label{sec:intro}
Are we alone in the universe? The question of whether there is life beyond Earth has occupied humanity for millennia. The  current time in human history presents an unprecedented opportunity to answer that question. Thousands of exoplanets have been discovered in the three decades since the early 1990s, revealing their ubiquity and diversity and that low-mass exoplanets, smaller than Neptune, are particularly common\footnote{\url{https://exoplanetarchive.ipac.caltech.edu}}. There are increasing numbers of potentially habitable rocky exoplanets being found orbiting nearby stars, while theoretical studies are also expanding the types of planets that could be habitable beyond Earth-like planets. At the same time, new large telescopes in space and on ground are beginning to attain the capability to detect atmospheric signatures of potentially habitable planets. These advancements in observations and theory are preparing the ground for the discovery and characterization of habitable exoplanets and possible biosignatures in their atmospheres. However, before being able to confidently identify both the locations and markers of life, we need to understand what constitutes habitability, what factors affect habitability, and what are the biosignatures expected from life in extraterrestrial habitable environments. 

Given the vast expanse of the universe and the outstanding diversity of planetary and exoplanetary  environments now known, it is hard to truly define the limits of habitability or the possible nature of life elsewhere. Nevertheless, at the present time, there is one place known to host life - the Earth. Therefore, our notions of habitability, alien life, and biosignatures are largely influenced by our experience on Earth, which serves as a convenient starting point. In the present chapter, we summarize a broad range of factors associated with habitability and biosignatures, but it is important to remember that there may still be environments and life elsewhere that lie beyond these considerations. In what follows, we first discuss the various factors affecting planetary habitability and different types of habitable environments. We then discuss considerations for identifying biosignatures, examples of terrestrial biosignatures and their applicability to exoplanets, the current state of the field and future prospects. 

\section{Habitability} \label{sec:habitability}
Habitability refers to the condition of an environment which satisfies the essential requirements for sustaining life. Conventional notions of habitability are motivated by requirements of life as found on Earth. A habitable environment is expected to meet four key requirements: (1) liquid water, (2) bio-essential elements, (3) energy source, and (4) environmental conditions, such as pressure and temperature, within certain limits. In what follows, we first provide an overview of these factors as known on Earth, and then discuss various other factors that can influence planetary habitability. Extensive reviews on exoplanetary habitability are presented in the following works: \citet{cockell2016habitability,cockell2024sustained, Kaltenegger2017, meadows2018exoplanet, schulze2018, kane_book2021, catlingbook2017}.

\subsection{Habitable Zone and Liquid Water} 
\label{sec:hz}

\begin{figure}
\begin{center}
\includegraphics[angle=0,width=\textwidth]{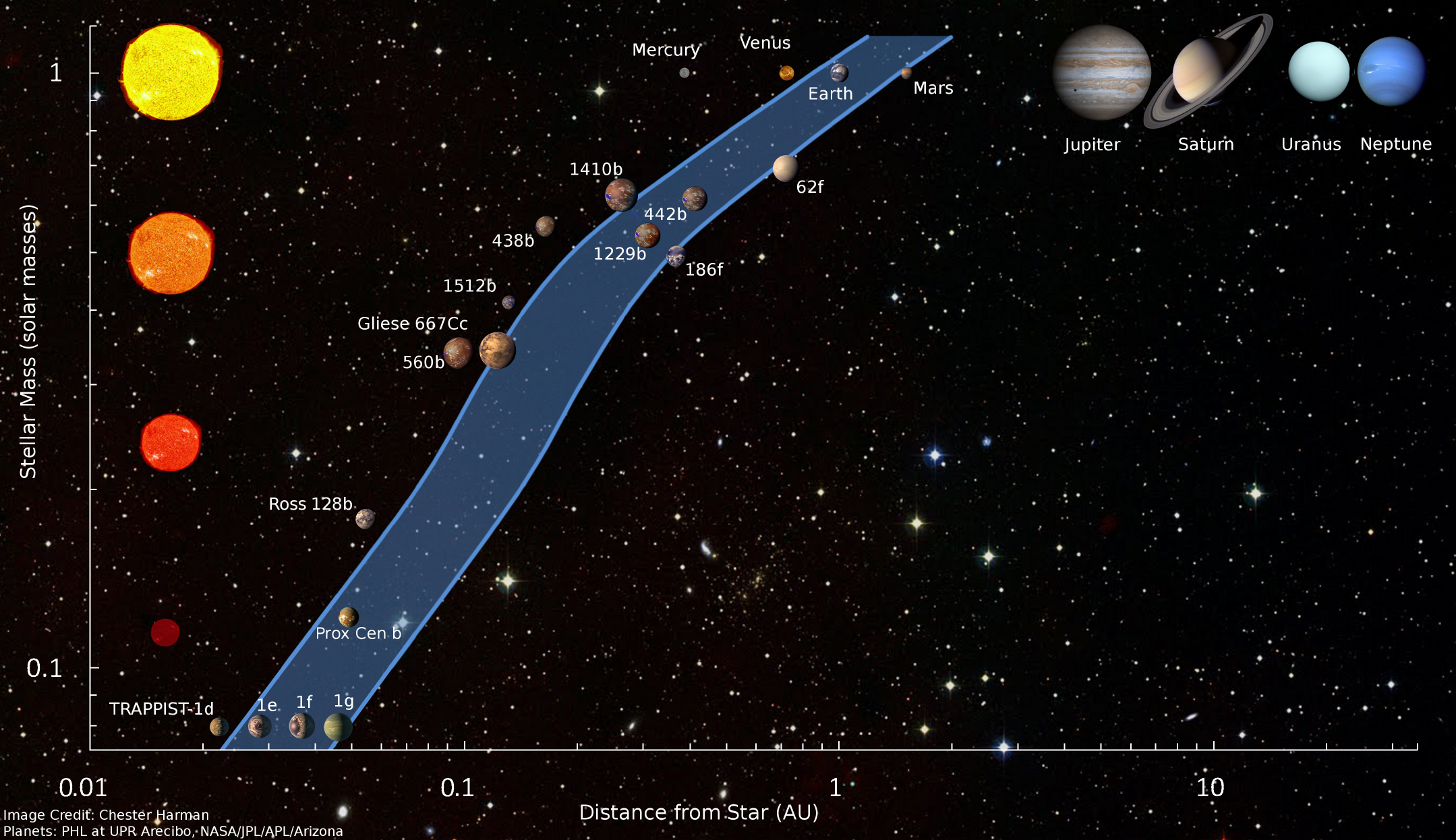}
\caption{Schematic of a canonical habitable zone for terrestrial planets around main sequence stars. The inner and outer edges of the HZ, shown in blue, correspond to the moist greenhouse limit and the maximum greenhouse limit, respectively \citep{kasting1993habitable, Kopparapu2013}. The solar system planets and several exoplanets are shown for reference. Image Credits: Chester Harman.}
\label{fig:hz}
\end{center}
\end{figure}

A nominal starting point to assess the habitability of a planet is its potential to host surface liquid water. Traditionally, the habitable zone (HZ) is defined as the region around a host star where an Earth-like planet can sustain liquid water on its surface \citep{kasting1993habitable}. Such a definition depends on a number of assumptions such as the atmospheric composition and albedo, internal heat, age and other properties of the planet \citep{meadows2018factors, selsis2007habitable}. In the canonical model for HZ estimates, an Earth-like planet is assumed with CO$_2$, H$_2$O and N$_2$ as prominent gases in the atmosphere \citep{kasting1993habitable, selsis2007habitable, Kopparapu2013}. The dominant greenhouse gas at the inner edge of the habitable zone (IHZ) is H$_2$O whereas that at the outer edge of the habitable zone (OHZ) is CO$_2$. A schematic diagram of the canonical habitable zone for terrestrial planets orbiting main sequence stars is shown in Figure~\ref{fig:hz}. The orbital separation of the IHZ typically corresponds to the moist greenhouse limit at which the surface temperature is high enough that the water vapour content in the stratosphere increases dramatically, which can cause significant photolysis of water and subsequent loss due to atmospheric escape. Another IHZ may be defined for the runaway greenhouse limit at which the oceans can evaporate entirely. The increased radiation closer to the star causes increased surface temperature, which in turn increases the water vapour content and greenhouse warming in the atmosphere. The OHZ corresponds to the maximum greenhouse limit, i.e. the distance at which the maximum greenhouse warming due to CO$_2$ is possible in the atmosphere. Beyond the OHZ, CO$_2$ in the atmosphere is no longer able to provide adequate warming and eventually condenses out leading to a substantially decreased greenhouse effect and the freezing of surface water. In this framework, the typical HZ around a sun-like star at the present day spans orbital separations of 0.99 - 1.70 astronomical unit (au), which includes the current orbital separations of Earth and Mars \citep{Kopparapu2013}; the moist greenhouse limit is estimated at 0.99 au and the runaway greenhouse limit at 0.97 au. For larger, and hotter stars, the HZ lies farther out. Conversely, for smaller, and hence cooler, stars the HZ can be significantly closer to the star. For example, as shown in Figure~\ref{fig:hz}, for a late M dwarf with a mass of 0.1 M$_\odot$ the HZ lies within 0.1 au, whereas for a typical F star of mass 1.2 M$_\odot$ the HZ can extend up to 2 au. 

While the canonical HZ serves as a useful guide, it is important to note that the extent of the HZ depends strongly on the assumed atmospheric, surface and interior properties of the planet.  For example, a significantly higher or lower albedo can move the inner edge of the HZ significantly closer to or farther from the star, respectively \citep[e.g.][]{selsis2007habitable}. Similarly, a significantly different atmospheric composition to Earth could cause different levels of greenhouse warming of the surface from that experienced on Earth. A more efficient and volatile greenhouse gas like molecular hydrogen (H$_2$) allows the OHZ to be significantly farther from the star \citep{Pierrehumbert2011,  madhusudhan2021habitability}. The amount of water vapour present in the atmosphere and contributing to the greenhouse warming also depends on the water inventory available on the surface. As such dry rocky planets may be expected to have closer Inner HZ boundaries compared to planets with higher water content. Another important factor to consider is the evolutionary stage of the system. The stellar luminosity varies over the age of the star thereby varying the HZ boundaries for a given planet over time. The Continuous Habitable Zone (CHZ) is defined as the distance from the star where an Earth-like planet would be able to sustain life on some part of the planet over geological time scales, e.g. billions of years (Gyr) \citep{rushby2013, cockell2016habitability}. 

\subsection{Bioessential Elements}
\label{sec:elements}
Life requires various elements that constitute molecules which act as the building blocks of biological structures and play a key role in various biochemical functions \citep{egli2009, cockell2016habitability}. Such bioessential elements include the major elements carbon, hydrogen, nitrogen, oxygen, phosphorus, and sulphur (also referred to as CHNOPS), as well as a number of minor elements which are present in trace quantities such as Fe, Na, K, Mg, Mn, Ni, Ca, Cu, Zn, Co, and Mo. Each of these elements serves a range of roles across the diversity of life we see on Earth, from the cellular to macroscopic levels. All life on Earth is carbon-based. C is the central element in all biomolecules thanks to its ability to readily form covalent bonds with other atoms resulting in a vast range of large organic structures essential for life, including DNA, RNA, proteins, lipids, carbohydrates and others. Hydrogen is another major constituent of biomolecules, thanks to the carbon-hydrogen covalent bonds in organic molecules and its presence in H$_2$O. Similarly, O is a major bioessential element, not only because of the dependance of most terrestrial life on H$_2$O and atmospheric O$_2$ but also because of the presence of O in a large number of macromolecules, including carbohydrates and proteins. Finally, N is another key component of important biological structures such as the DNA and aminoacids, which are the building blocks of proteins, besides performing a wide range of other biological functions. 

Together, the four major elements C, H, O and N, constitute about 97\% by mass of living organisms on Earth and define most of their biological structure and function. Another bioessential element P forms a key constituent of the DNA and RNA besides being involved in various biological functions, including energy production through the ATP molecule, cell structure through phospholipids and other functions, primarily through phosphates. The remaining major element S is essential for important amino acids and enzymes and as an energy source for several microorganisms. It is important to note that the roles of the major elements discussed here are but representative examples from a vast range of functions they perform across the complex biosphere on Earth. Besides the major elements, life on Earth also uses a large range of other minor elements for various biological functions including structure, growth, regulation, and energy requirements. Therefore, the availability of such elements in adequate quantities is considered to be an important requirement for life on a planet. 

\subsection{Energy}
\label{sec:energy}
All life needs energy for survival and biological functions such as maintenance, growth, movement and replication \citep{schulze2018, cockell2016habitability}. On Earth, the primary producers (autotrophs) derive energy from two primary sources, sunlight and chemical reactions, to convert simple molecules into complex organic compounds which store energy. The consumers ('heterotrophs') derive energy from the autotrophs or other heterotrophs. The process of deriving energy from sunlight is known as phototrophy and that from chemical reactions is known as chemotrophy. The corresponding organisms are referred to as phototrophs and chemotrophs, respectively. There are also organisms which can use both mechanisms. In chemotrophs, the energy is derived from redox reactions which involve electron transfer from a reducing agent to an oxidising agent. Depending on whether the electron donor is an organic or inorganic molecule the organism is classified as a chemoorganotroph or chemolithotroph, respectively. These mechanisms all involve converting the available energy to chemical energy in the form of organic molecules, particularly adenosine triphosphate (ATP), which can store, transport and release energy as needed. The ATP molecule, often described as the ``energy currency" in a cell, plays a central role in energy transport between cells and is produced in all living organisms.

Chemotrophs on Earth use energy from a wide range of redox reactions. The electron donor, i.e. the reducing agent, in the redox reaction provides free electrons which are transferred across a series of compounds, mainly a chain of proteins in the cell membrane. This, in turn, causes positive ions, e.g. protons, to flow across the membrane, creating a proton gradient which provides the required electrochemical potential energy for the synthesis of ATP. Numerous compounds are used by chemotrophic life as electron donors, from simple molecules such as H$_2$ to complex hydrocarbons as well as metal ions such as Fe$^{2+}$. One of the most primitive forms of chemotrophic life, methanogens, generate energy from the redox reaction involving CO$_2$ and H$_2$ to result in CH$_4$ and H$_2$O, a process known as methanogenesis. Here H$_2$ acts as a very efficient electron donor in the reduction of CO$_2$. This mechanism has been a significant source of CH$_4$ in the Earth's atmosphere since the early epochs. 

Phototrophs use energy from sunlight and available sources of carbon, such as CO$_2$, to create organic compounds such as carbohydrates and adenosine triphosphate (ATP). A prime example of phototrophy is photosynthesis in plants, whereby interaction between chlorophyll and sunlight in the presence of H$_2$O result in O$_2$ and ATP, and subsequent capture of CO$_2$. The molecules created in such processes, such as carbohydrates, can also be later oxidised e.g. through O$_2$ respiration in organisms, to release energy for biological needs. Besides plants, a wide range of microorganisms use photosynthetic mechanisms to use sun-light to meet their energy requirements. While oxygenic photosynthesis is the most common in phototrophs, some microorganisms also use reactions involving sulphur-based compounds, resulting in the release of sulphur instead of O$_2$ \citep{schulze2018}. 

\subsection{Environmental Conditions}
\label{sec:environment}
Life on Earth is known to be ubiquitous and resilient, with organisms known to survive under extreme conditions. Nevertheless, organisms are known to be sensitive to several environmental parameters which can affect their survival and key biological functions \citep{Rothschild2001,merino2019}. These parameters include: (a) temperature, (b) pressure, (c) pH, (d) salinity, and (c) radiation. 

Temperature is arguably the most important environmental parameter as it determines the phase of water at a given pressure and also affects the biological functions and structural integrity of an organism. Water is in liquid phase between 0-100 $^\circ$C (273-373 K) for standard pressure of 1 bar, and over a broader range of temperatures for higher pressures. While most life on Earth is adapted to function normally over a narrow temperature range, e.g. 0-40 $^\circ$C, extremophile organisms are known to survive at temperatures between -15 and 122 $^\circ$C (258-395 K). The low temperature limit corresponds to microorganisms found under antarctic ice whereas the high temperature limit corresponds to those found near hydrothermal vents and hot springs. Another upper limit on the temperature is governed by the fact that proteins can denature above $\sim$100$^\circ$ C and macromolecules become unstable beyond $\sim$150$^\circ$ C, which, therefore, sets a structural upper-limit for carbon-based life as found on Earth \citep{Rothschild2001,merino2019}. 

Similar to temperature, pressure is another important parameter which governs the phase of water and also affects the structure and function of organisms. While most life on Earth operates near standard pressure of 1 bar, organisms are known to survive at pressures up to 1000 bar, corresponding to the deepest parts of the oceans. In principle, water can remain in liquid form for even higher pressures, up to $\sim$60,000 bar for temperatures below 374 $^\circ$C (647 K). However, as with temperature discussed above, proteins disintegrate at pressures above a few thousand bars, setting a natural upper-limit for carbon-based Earth-like life. Other environmental conditions, including pH, salinity and UV radiation, also affect the survival and biological functions of organisms \citep{Rothschild2001,merino2019}. 

\subsection{Other Factors affecting Habitability} 
A number of other factors can affect the habitability of a planet. Here we discuss some of the major factors, which include a range of internal, atmospheric and surface properties of a planet but also, properties of its host star and the planetary system as a whole. 

\subsubsection{Atmospheric Escape }
\label{sec:escape}
The presence of an adequate atmosphere is essential to provide the greenhouse warming required to maintain liquid water on the surface of a planet. The extent and composition of a planetary atmosphere are strongly influenced by its formation and evolutionary history, in which atmospheric escape processes play a prominent role \citep{lammer2012escape}. A planet is typically formed with a primordial H$_2$-rich atmosphere, which reflects the composition of gas accreted from the protoplanetary disk along with trace quantities of other elements accreted through solids containing ices and refractory material. After formation, the initial atmosphere can be significantly eroded due to atmospheric escape over the planetary lifetime. For planets with large masses and thick primordial atmospheres, such as the ice giants and gas giants in the solar system, a significant fraction of the primordial atmospheres can still be retained. Their evolved atmospheres continue to be H$_2$-rich, and are referred to as primary atmospheres. However, for terrestrial-like planets, such as Earth, Venus and Mars, the primordial atmospheres can be largely eroded and their evolved atmospheres are mostly a result of outgassing from the planetary interior along with any remnant heavy elements that survived atmospheric escape. These are known as secondary atmospheres, and tend to be dominated by heavy molecules such as N$_2$ in the case of Earth and CO$_2$ in the case of Venus and Mars.  Planets of intermediate masses, e.g., sub-Neptunes, can host primary and/or secondary atmospheres depending on their specific conditions and evolutionary histories. No such planets exist in the solar system but they dominate the exoplanet population and their atmospheric compositions are presently not fully known. 

The degree of atmospheric escape and resulting atmospheric composition depend on other factors besides the planet mass, which can affect planetary habitability. These include the planet age, host star, orbital separation, magnetic field, and internal composition. In the initial stages of the planetary evolution the atmospheric escape is primarily caused by thermal escape due to extremely energetic (EUV and XUV) radiation from the young and active host star. For terrestrial planets in the solar system the primordial atmospheres are thought to have been largely eroded within a few hundred million years after formation, followed by the formation of secondary atmospheres. At later times, atmospheric escape is contributed by both thermal and non-thermal escape processes such as kinetic and electromagnetic interactions between particles in the stellar wind and the planetary atmosphere. The resulting atmospheric composition at a later time is governed by the balance between the atmospheric escape and replenishment from surface-atmospheric interactions, such as outgassing of key gases from the planetary interior. %outgassing of CO2, H2O, and N2.. how it worked for Earth, venus, mars..  

\subsubsection{Magnetic Fields} 
\label{sec:magnetic}
The presence of a planetary magnetic field is generally thought to inhibit atmospheric loss due to stellar winds. Several factors influence how, and to what extent, a magnetic field could protect a planetary atmosphere \citep{cockell2024sustained,gronoff2020}. One of the mechanisms of atmospheric escape in a planet is due to the interaction between the stellar wind and the planetary atmosphere. The stellar wind is composed of charged particles from the stellar corona that travel outward along the stellar magnetic field lines. In the absence of a planetary magnetic field the energetic stellar wind can significantly deplete the particles in the upper atmosphere of the planet. A planetary magnetic field deflects the stellar wind and prevents such loss. This is particularly relevant for thin planetary atmospheres of terrestrial planets, for which the gravity may not be strong enough to prevent escape. This is thought to be the case for Mars, whose atmosphere may have been significantly eroded due to its low gravity and the lack of a strong magnetic field. On the other hand, the magnetic field of the Earth is instrumental in protecting the atmosphere from the solar wind. However, recent studies are suggesting that a planetary magnetic field alone may not be a strong determinant of atmospheric protection or escape \citep{gronoff2020}; for example, Venus has a dense atmosphere without a strong magnetic field. 

It is also becoming more evident that rocky planets orbiting low-mass stars (M dwarfs) are more susceptible to atmospheric loss from stellar winds. A magnetic field is generated by a magnetic dynamo due to a conductive layer in the planetary interior. The Earth's rotating outer core, composed of liquid iron, provides the conducting layer for the dynamo that powers the magnetic field. Low-mass stars including M and K dwarf stars are expected to be highly active leading to strong star-planet interactions. Secondly, for planets that are tidally locked, the magnetic field is expected to be weaker due to a slowly rotating internal dynamo. Thus, tidally-locked habitable rocky planets orbiting M dwarfs are more susceptible to atmospheric loss in the absence of a magnetic field \citep{cockell2024sustained}. However, even the presence of a magnetic field may not fully prevent atmospheric loss due to stellar winds for such systems \citep{gronoff2020}. 

\subsubsection{Geological activity and Geochemical Cycling} 
\label{sec:geology}
Geological activity and geochemical cycling of nutrients can have a significant effect on the atmospheric composition of a planet and, hence, on its habitability to some extent \citep{Kaltenegger2017, meadows2018factors, cockell2016habitability}. On Earth, geological activity is manifested in the form of plate tectonics and volcanism, which have important consequences for the formation of surface topography, availability of bioessential elements, atmospheric composition, and interior processes, including magnetic field generation. The Earth's outermost solid layer (the lithosphere), which includes the crust and upper mantle, is composed of several (7-8) large separate segments (plates) which form the basis of plate tectonics. The relative movements and interactions between these plates result in topographic features and geological activity on the surface, such as mountains, ridges, and volcanoes. Volcanic activity, both from plate tectonics as well as direct eruption through the crust, results in outgassing of prominent greenhouse gases such as H$_2$O, CO$_2$ and SO$_2$ from the interior into the atmosphere. Outgassing of such gases is important to maintain the greenhouse warming required for a habitable surface temperature, especially considering their possible depletion through atmospheric escape processes. Plate tectonics on Earth is also thought to influence the magnetic dynamo in the interior thereby affecting the magnetic field which, in turn, protects the atmosphere from the solar wind as discussed above. Earth is the only planet known to host plate tectonics. The other terrestrial planets, e.g. Venus and Mars, contain a `stagnant lid' with a single plate. While geological activity has been beneficial for habitability on Earth, it is not known if planet tectonics is essential for habitability on all planets in general. 

On Earth, geological activity also plays an important role in the geochemical cycling of bioessential elements as well as regulation of the climate. As discussed in section~\ref{sec:elements}, life on Earth requires several bioessential elements, including the major elements (CHNOPS) as well as several minor elements. These elements are made available through a wide range of biogeochemical cycles which represent an intricate interplay between the ocean, atmosphere, land and biosphere, through climatic, geological, and biological processes. In particular, weathering of mineral-rich rocks in contact with various forms of water is a prominent source for several bioessential elements in the ocean. 

At the same time, geological activity also influences the atmospheric composition and, hence, the climate and surface conditions of the planet. For example, the geochemical cycling of CO$_2$ is a key component of the carbon-cycle and a major regulator of the surface temperature of Earth over geological timescales, through the carbonate-silicate cycle. In this process, CO$_2$ in the atmosphere reacts with rain water to form carbonic acid which causes chemical weathering of silicate-rich rocks to release metal ions and bicarbonate ions. The metal and bicarbonate ions are carried to the ocean where marine organisms convert them into carbonates, e.g. calcium carbonate in the form of bones and shells, and are eventually deposited on the ocean floor. Over time these deposits are carried deeper into the mantle through plate tectonics, where the high temperatures and pressures convert the carbonates back to silicate minerals and release CO$_2$. The CO$_2$ is released back to the atmosphere through volcanic emissions and other avenues, completing the cycle. An increase in the atmospheric CO$_2$ increases the greenhouse warming, and, hence, the surface temperature and evaporation, eventually causing more rain and weathering. This, in turn, removes more CO$_2$ from the atmosphere, thereby cooling the surface, and deposits it in the ocean. This cycle operates on roughly million-year timescales, and acts as a natural thermostat for the Earth's surface temperature on geological timescales. Similar or other geochemical cycles may be needed on other rocky exoplanets to maintain habitable conditions on long timescales. 

\subsubsection{Host Star} 

Among all the external factors that can influence the habitability of a planet, the nature of the host star is most important, along with the orbital separation from the star \citep{meadows2018factors, cockell2016habitability, kane_book2021}. The stellar radiation plays a major role in determining the atmospheric and surface conditions while also being the primary source of energy for most life as known on Earth. For main sequence stars, the luminosity and effective temperature ($T_{\rm eff}$) of the star vary with mass; more massive stars are more luminous, larger and hotter. The luminosity represents the total energy output of the star per unit time, and depends on the effective temperature and size of the star. The effective temperature is also represented by the stellar spectral type, ranging from over 7000 K for F stars and 5000-6000 K for sun-like G stars to 2500 K for the coolest M stars, also known as M dwarfs. Both the luminosity and spectral type influence the atmospheric and surface conditions of a planet. As discussed in section~\ref{sec:hz}, the mass of the star determines the extent of the HZ, which is closer in for the less luminous low-mass stars compared to sun-like stars. The mass also determines the nature of the radiation, i.e. the spectral energy distribution (SED), as the SEDs from cooler stars peak at redder wavelengths. Whereas the solar spectrum peaks at visible wavelengths, the spectrum of a 3000 K M dwarf peaks in the near infrared, around 1 $\mu$m. This in turn affects the atmospheric temperature structure, photochemistry, and overall climate, as well as the surface conditions and nature of life. While life on Earth is adapted to utilize solar radiation in the visible as the primary source of energy, life on a planet around an M dwarf may evolve to survive predominantly on near-infrared radiation under different atmospheric and surface conditions.

The stellar radiation can also have adverse effects on planetary habitability and influence the long term evolution of the atmosphere. As discussed in section~\ref{sec:escape}, high-energy EUV and XUV radiation from the star drive atmospheric thermal escape whereas strong stellar winds due to stellar magnetic fields can drive non-thermal escape, sculpting the planetary atmosphere. High-energy radiation is also detrimental to life as at high intensity it can damage biomolecules at a cellular level. In addition, strong flares from active stars are further detrimental to the atmosphere and life. Stellar activity tends to be stronger for lower mass stars due to their more convective interiors which facilitate stronger magnetic dynamos, and for younger stars which have faster rotation periods. On the other hand, lower mass stars are more favorable for detecting habitable planets, due to their closer HZ boundaries and higher planet-star size ratio. Overall, adequate atmospheric protection due to gravity and/or a planetary magnetic field are required to preserve the planetary atmosphere and the biosphere. Finally, the age of the star also plays an important role on the planetary atmosphere. As alluded to above, stars are more active when younger and therefore induce stronger atmospheric loss early on, which sets the initial conditions for subsequent planetary evolution.

\subsubsection{Dynamical Properties} % Planetary system dynamics} 
The prospects for habitability of a planet in the HZ are also affected by a wide range of dynamical properties, both of the planet as well as the planetary system as a whole \citep{meadows2018factors, cockell2016habitability, kane_book2021}. The planetary properties include its rotation period, orbital eccentricity, obliquity and spin precession, all of which affect its atmospheric dynamics, climatic states, long term atmospheric evolution, and its surface topology. For example, the obliquity of the Earth, i.e. the tilt of its spin axis relative to its orbital plane, is responsible for the seasons, and the combination of different dynamical properties affected the long-term evolution of its climate. Finally, the presence of other bodies in the system, e.g. planets and/or satellites, also affect the dynamical stability of the planet and its interactions with the environment, such as giant impacts which can be disruptive for life. The presence of nearby planets may reduce the frequency of impacts by scattering away the impactors, as Jupiter is considered to have done for the Earth, or increase the frequency of impacts, depending on the orbital architecture of the planetary system. 

\subsection{Types of Habitable Environments} \label{sec:environments}

Earth is the only planet currently known to be inhabited. However, numerous efforts are currently underway to search for extraterrestrial habitable environments, both in the solar system and in exoplanetary systems. Here we briefly outline such objects and their prospects. 

\subsubsection{Habitability in the Solar System}
As discussed in section~\ref{sec:hz}, the nominal habitable zone (HZ) around the Sun extends between 0.99-1.7 au, assuming an Earth-like atmosphere and surface conditions \citep{kasting1993habitable,Kopparapu2013}. Earth at 1 au from the Sun lies close to the inner edge of the nominal HZ \citep{selsis2007habitable, yang2013stabilizing}. A higher albedo or lower greenhouse warming compared to Earth can move the HZ inward while a lower albedo or higher greenhouse warming can move it outward \citep{selsis2007habitable, yang2013stabilizing}. The search for life beyond Earth in the solar system is focused primarily on Mars and icy moons of the giant planets.

{\it Mars:} The only other planet in the solar system within the HZ is Mars, at an orbital separation of 1.5 au. However, Mars hosts a significantly thinner atmosphere compared to Earth, with an average surface pressure of 6 mbar. The atmospheric composition comprises of CO$_2$ (95\% by volume), N$_2$ (3\%), Ar (2 \%), and other trace gases. The greenhouse warming caused by the Martian atmosphere is insufficient to permanently sustain substantial liquid water on its surface, though there is evidence for frozen water and potentially sub-surface liquid water \citep{lauro2021}. Even though there is no concrete evidence for life on Mars at present, it is possible that the planet may have been habitable in the past with the possible presence of liquid water due to a thicker atmosphere. 

{\it Icy Moons:} Other potential habitable environments in the solar system are the icy moons of giant planets, departing from the conventional definition of the HZ. All the giant planets are outside the HZ, with Jupiter at 5 au being the nearest to the Sun. Giant planets are also considered non-habitable owing to their thick H$_2$-rich atmospheres which lead to high pressures and temperatures that preclude a liquid water surface. However, some of the satellites of the giant planets are capable of hosting sub-surface liquid water oceans which could be conducive for habitability. Some examples include Jupiter's icy moons Europa, Ganymede and Callisto, and Saturn's moon Enceladus. Their low surface temperatures, due to the low solar radiation received and lack of significant atmospheres, cause a crust of water ice at the surface. However, internal heat from tidal interactions can lead to a layer of liquid water below the icy crust. The deeper interior below the water layer can be composed of rock or high-pressure ice, which in turn affect the availability of bioessential elements and metal ions required for biochemical requirements of life in such environments. 

{\it Other Environments:} More exotic environments have also been considered for habitability in the solar system. On one end is Saturn's moon Titan, which is believed to host liquid hydrocarbons, e.g. methane lakes, on the ultra-cold (-180 $^o$C) surface of an icy and rocky interior and an N$_2$-rich atmosphere. Despite the lack of liquid water under such conditions, investigations focus on the possibility of alternate biota that can survive in other solvents such as hydrocarbons. At the other extreme in temperature is Venus, with a volcanically active scorched surface (465 $^o$C) devoid of liquid water and clouds composed of sulphuric acid. However, studies have suggested the potential habitability of aerial microbial life in the high-altitude clouds of Venus, where temperatures decrease to those comparable to Earth-like environments. All these avenues present several open questions in all aspects of habitability, including the requirement of liquid water, the availability of bioessential elements, the energy sources, and the limits on environmental conditions that can host life. 

\subsubsection{Habitability in Exoplanetary Systems}
\label{sec:habitable_environments}

Exoplanetary systems greatly expand the range of possible habitable environments in the search for life elsewhere. This is due to both the high occurrence rate of exoplanets and the wide range of stars they orbit. The exoplanet host stars span a wide range, from sun-like and larger stars to low-mass M dwarf stars that can be ten times smaller than the Sun, implying a wide range of orbital separations for the HZ depending on the stellar type. The exoplanets also span a wide range of masses and radii which allow for diverse internal structures and atmospheric compositions that may be conducive for habitability. 

{\it Rocky exoplanets:} A natural goal in the search for habitable planets beyond the solar system is an Earth analog, i.e. an Earth-like rocky planet in the habitable-zone around a sun-like star. More generally, a rocky planet is defined as a planet whose interior is composed primarily of rocky and metallic material, e.g. a combination of silicate-based rocks and iron. For reference, the Earth has nearly two-thirds its mass in silicates and one-third in iron, with a small fraction ($<$0.1\%) in volatile compounds including water and atmospheric gases. An exact Earth-analogue has not yet been discovered. However, tens of potentially habitable rocky exoplanets have been discovered, primarily around low-mass stars for which the HZ is at shorter orbital distances where planets are easier to detect compared to sun-like stars. The estimates of their radii and masses range between $\sim$1-2 R$_\oplus$ and $\sim$1-10 M$_\oplus$, respectively, spanning exo-Earths and super-Earths. A diverse range of interior and atmospheric compositions are theoretically possible for such planets. 

For habitability of exo-Earths and super-Earths, greenhouse warming may be contributed to by diverse atmospheric compositions, including prominent volatiles such as N$_2$, CO$_2$, H$_2$O, O$_2$, CH$_4$ and/or H$_2$ in varied proportions. The HZ boundaries, which typically assume Earth-like atmospheres, can vary significantly with atmospheric composition. For example, the outer habitable boundary can be much farther out for H$_2$-rich atmospheres compared to Earth-like or CO$_2$-rich atmospheres, as H$_2$ can be an efficient greenhouse gas and has a lower condensation temperature than CO$_2$ or H$_2$O. No atmospheric chemical detections have been made for a habitable rocky exoplanet to date. However, several known HZ exoplanets are considered to be good targets for atmospheric observations, including the transiting exoplanet system TRAPPIST-1 \citep{gillon2017seven} and the nearest, non-transiting, exoplanet Proxima Cen b \citep{anglada2016}, both of which orbit small and cool M dwarf stars. While such stars are the best targets for detecting habitable planets, they can also be very active and, thus, detrimental for atmospheric retention and habitability, especially for rocky planets with thin atmospheres.

{\it Water worlds:} These are planets whose interiors contain a large fraction of water. Despite the ocean covering $\sim$70\% of the Earth's surface the mass fraction of water in the Earth is $<$0.1\%. By contrast, water worlds can have tens of percent of water by mass. A habitable water world is characterised by a habitable ocean-covered surface with a deep layer of liquid water which can transition into high-pressure ice at depth followed by a rocky interior deeper down. The atmospheres of such planets may also be varied, e.g. dominated by N$_2$, CO$_2$, H$_2$O, O$_2$ and/or H$_2$, as for the rocky planets. Water worlds are further classified into ocean worlds and hycean worlds depending on the atmospheric composition. Ocean worlds are water worlds with terrestrial-like secondary atmospheres comprising predominantly of heavy molecules such as N$_2$, CO$_2$, H$_2$O and/or O$_2$ \citep{leger2004}. On the other hand, hycean worlds are water worlds with atmospheres dominated by H$_2$; the word ``Hycean" is a portmanteau of ``Hydrogen" and ``Ocean" \citep{madhusudhan2021habitability}. While water worlds meet several of the basic requirements for life -- of liquid water, available energy and appropriate thermodynamic conditions -- open questions remain on whether they are conducive for the origin and sustenance of life, and what albedos are needed to maintain the required atmospheric temperatures and liquid water on the surface. An important question is whether adequate bioessential elements would be available in the required concentrations considering that on Earth such elements are obtained through interactions between the water and mineral-rich rocks which may not be accessible on water worlds. Several avenues have been suggested that may overcome this limitation \citep{madhusudhan2023chemical}. 

A central advantage of water worlds is their observability. For a given planet mass, water worlds are less dense and hence larger in size compared to rocky planets. Their radii span $\sim$1-2.6 R$_\oplus$ for masses of 1-10 M$_\oplus$ \citep{madhusudhan2021habitability}. The larger sizes make them more detectable in exoplanet transit surveys. More specifically, for hycean worlds, their H$_2$-rich atmospheres make them even more favourable for atmospheric observations compared to both rocky planets and ocean worlds with other atmospheric compositions. H$_2$ being the lightest molecule possible results in a lower atmospheric mean molecular weight (MMW) and, hence, larger scale height and larger atmospheric absorption, compared to other compositions. At the same time, the greenhouse warming due to the H$_2$-rich atmosphere also significantly expands the HZ, as also discussed above for rocky planets, thereby increasing the sample of potentially habitable candidates for atmospheric observations.  Several exoplanets have been suggested as promising candidates for habitable water worlds, specifically candidate hycean worlds. Recent observations with the James Webb Space Telescope (JWST) led to the first detection of carbon-bearing molecules on a candidate hycean world, K2-18~b \citep{madhusudhan2023carbon}.  % \citep{madhu2023_JWST}%

\section{Biosignatures} \label{sec:biosig}

The search for life in a habitable environment involves the detection and identification of a biosignature. A biosignature is any observable aspect of a habitable environment which indicates the presence of life. This can be a single attribute, such as a gaseous molecule, or a combination of attributes which together provide robust evidence for life. Earth is the only planet currently known to host life. Therefore, notions of biosignatures are typically motivated based on indicators of life on Earth. The term `biosignature' is sometimes used interchangeably with  biomarker, and usually refers to a single molecule of biological origin. Here we discuss biosignatures expected in habitable environments, with the Earth as a canonical example, followed by current status and future prospects for biosignature detection on exoplanets. Extensive reviews on biosignatures can be found in the following works: \citet{catlingdavid2018exoplanet, schwieterman2018exoplanet, schwieterman2024overview, seager2016toward}.

\subsection{Atmospheric Biosignatures}
The most accessible biosignatures are biogenic gases in a planetary atmosphere inferred through spectroscopic observations. For an exoplanet, the observations can be of a transmission spectrum, emission spectrum or a reflection spectrum \citep{Madhusudhan2019}. On Earth, life produces a wide range of molecules that are released into the atmosphere at varied concentrations. However, several of these gases can also be produced by various abiotic processes, including geochemical and atmospheric processes which can also be prevalent in extraterrestrial environments. On the other hand, many of the gases are produced in too small quantities for them to be meaningfully detectable through astronomical observations. Therefore, it is important to consider what constitutes an ideal biosignature \citep[e.g.][]{seager2012astrophysical, seager2016toward, catlingdavid2018exoplanet, schwieterman2018exoplanet}.

An ideal biosignature in a given habitable environment is typically expected to have the following 
properties: 
\begin{itemize}
\item{It has no significant abiotic sources in the environment considered, either atmospheric or geochemical.}
\item{It has strong spectral features that are observable.}
\item{It can be produced by life in significant enough quantities to be detectable.}
\end{itemize}

The most prominent gases in the Earth's atmosphere that satisfy these criteria are oxygen (O$_2$), and its derivative ozone (O$_3$), and nitrous oxide (N$_2$O). Another gas that satisfies most of these criteria is methane (CH$_4$), though it is also produced geochemically, albeit in small quantities. However, when considering an extraterrestrial habitable environment, the reliability of these gases as biosignatures depends on the specific environmental context. For example, it has been shown that O$_2$ can be produced photochemically in atmospheres of terrestrial planets orbiting M dwarfs under certain conditions \citep{meadows2018exoplanet}. On the other hand, while many of the trace molecules may not be detectable in an Earth-like exoplanetary atmosphere, they may still be detectable for H$_2$-rich atmospheres where the spectral features are amplified due to the low mean molecular weight and large scale height of the atmosphere \citep{seager2013b_h2, madhusudhan2021habitability,schwieterman2024overview}. Therefore, the assessment of an ideal biosignature also needs to account for its observability in the specific environment considered. In what follows, we discuss the prominent gaseous molecules that have been proposed as good potential biosignatures in habitable environments. 

It is instructive to consider the prominent gases produced by life on Earth and their characteristics as potential biosignatures. The spectral features of several notable molecules are shown in Figure~\ref{fig:spectra}. An extensive literature on such gases in discussed in several recent reviews \citep{schwieterman2018exoplanet, schwieterman2024overview, seager2016toward}. Such gases may be considered under two broad categories based on their metabolic origins \citep{seager2012astrophysical}: (a) primary metabolic byproducts, and (b) secondary metabolic byproducts. Primary metabolic byproducts are gases that are released by organisms as a result of basic 
metabolic functions such as energy generation and survival. These include gases such as O$_2$ generated from photosynthesis, CH$_4$ generated by methanogenic bacteria and CO$_2$ released by animals. Several of these gases are produced in significant quantities in the Earth's biosphere, and some of them also have significant abiotic sources, such as CO$_2$ released from volcanoes. On the other hand, secondary metabolic byproducts are gases that are produced during specific functions of organisms, e.g. those related to stress or fitness. These gases are produced in small quantities but are exclusively released by life with no significant abiotic sources on Earth. Examples of such gases include Dimethyl Sulphide (DMS), Methyl Chloride and Methyl Mercaptan. In what follows, we discuss prominent gaseous molecules that have been proposed as good potential biosignatures to search for in exoplanetary environments motivated by their presence in the Earth's atmosphere. 

\begin{figure}
\begin{center}
\includegraphics[angle=0,width=0.5\textwidth]{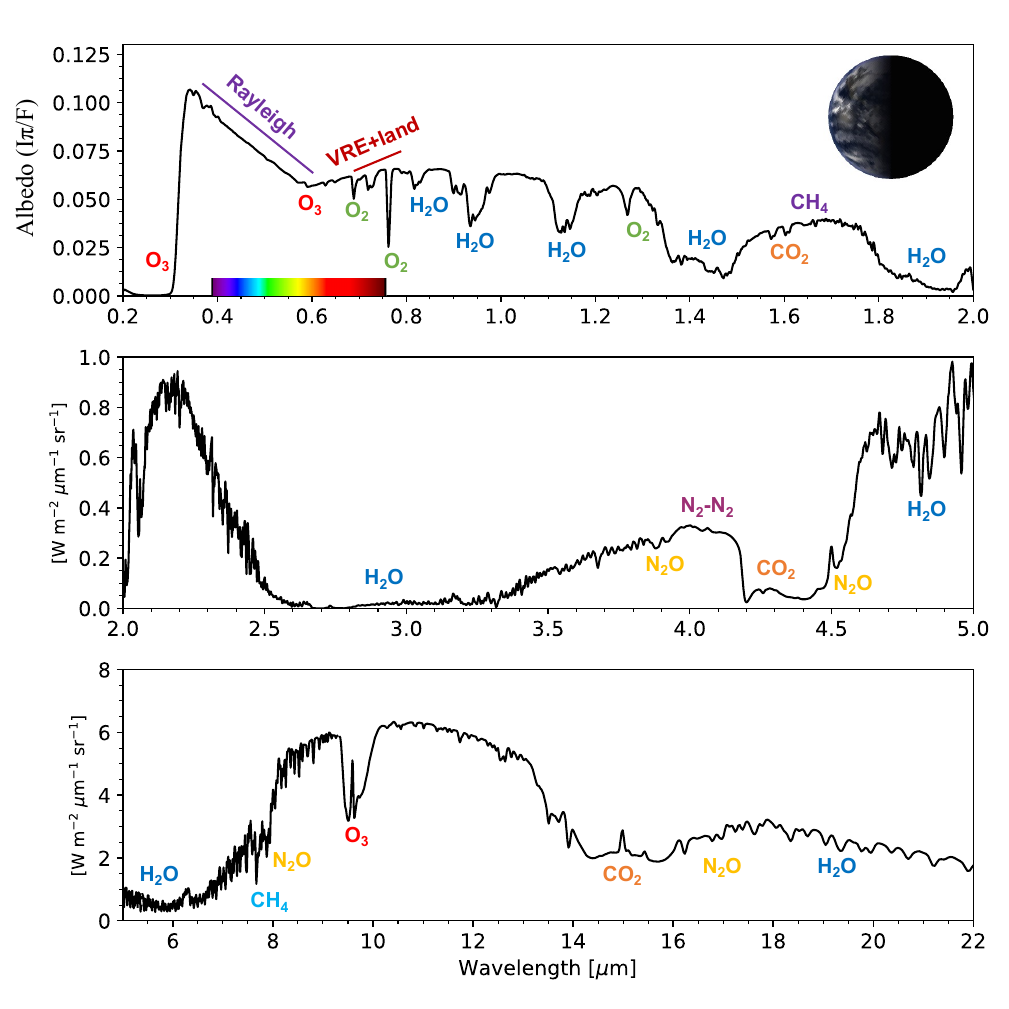}
\caption{Simulated spectrum of the Earth at half illumination showing notable spectral features including those from several biosignature gases \citep[from][]{schwieterman2024overview}. The top panel shows the apparent albedo spectrum due to reflected light between the UV and NIR (0.2-2 $\mu$m), the middle panel shows spectral radiance between 2-5 $\mu$m which includes both reflected light and thermal emission, and the bottom panel shows the spectral radiance in the mid-infrared which is primarily due to thermal emission.}
\label{fig:spectra}
\end{center}
\end{figure}

\begin{figure}
\begin{center}
\includegraphics[angle=0,width=\textwidth]{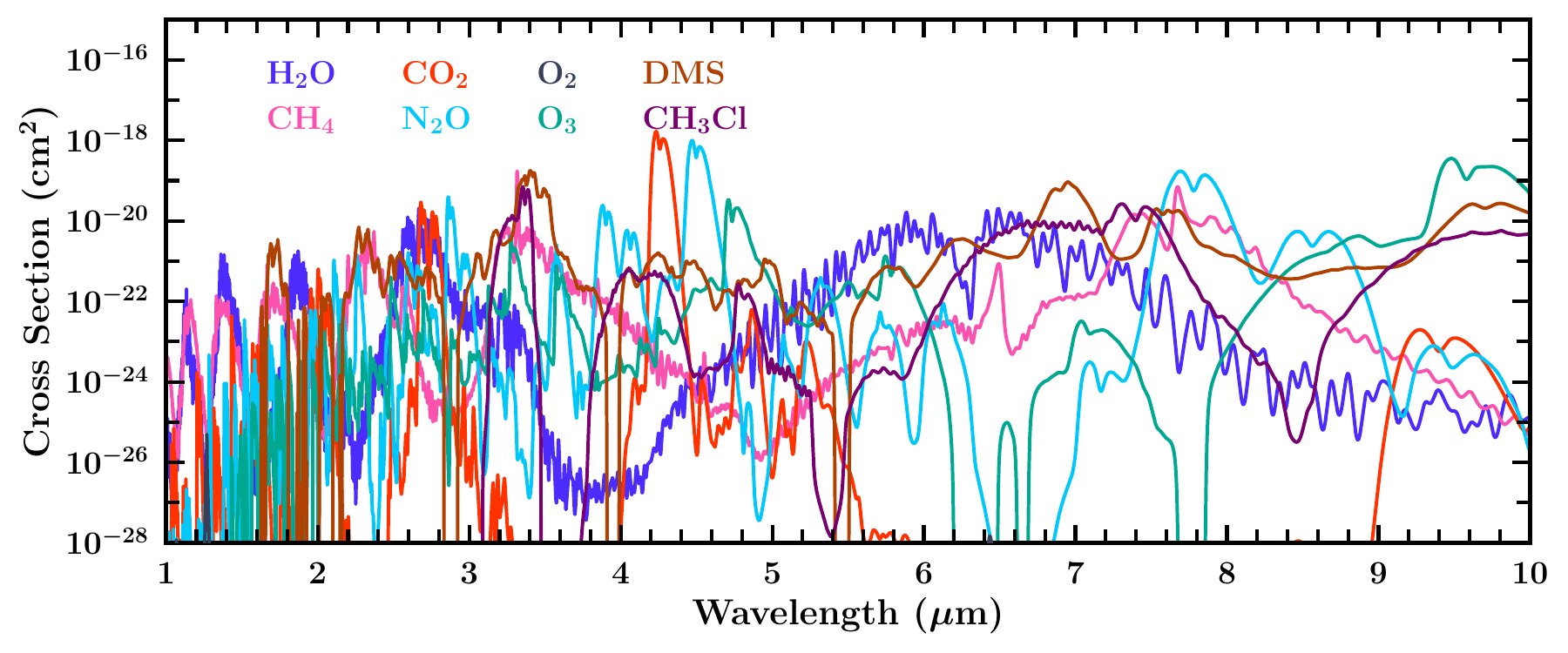}
\caption{Absorption cross sections of notable molecules. H$_2$O and CO$_2$ are prominent greenhouse gases in the Earth's atmosphere. They are also produced by life but are not considered as biosignatures due to their significant abiotic sources. The remaining molecules shown are predominantly produced by life on Earth, though some of them, such as CH$_4$, also have some abiotic sources; see section~\ref{sec:biosig}. All the molecules have strong absorption cross sections in the infrared but their abundances in the atmosphere can vary widely depending on the specific environmental context. The cross sections for H$_2$O, CH$_4$ and CO$_2$ are obtained following \citep{gandhi2020} for a pressure of 1 bar and temperature of 300 K, based on the ExoMol \citep{ExoMol2016} and HITRAN \citep{HITRAN2016} databases, and the rest from the HITRAN database.}
\label{fig:spectra}
\end{center}
\end{figure}

\subsubsection{Prominent Biosignature Gases}

The ideal goal in the search for life elsewhere is to detect a biosignature in an Earth-like atmosphere which is a known benchmark for habitability. However, a diversity of atmospheric compositions are possible for habitable rocky planets and ocean worlds with terrestrial-like secondary atmospheres. Such atmospheres may be dominated by a combination of N$_2$, O$_2$, CO$_2$ and/or H$_2$O, along with other trace species. Considering an Earth-like biosphere, the most prominent molecules traditionally considered as potential biosignatures are O$_2$, O$_3$, N$_2$O and CH$_4$. While life on Earth produces other gases in significant quantities, such as CO$_2$ and H$_2$O, they are readily available from natural sources and hence less relevant as biosignatures on their own. On the other hand, a number of other gases produced by life in smaller quantities could be important biosignatures as discussed below. It is also becoming increasingly evident that no single molecule may qualify as a robust biosignature and, instead, combinations of molecules may be required. Here, we discuss prominent gases in the Earth's atmosphere which are known to be produced by life and have been proposed as good biosignatures in extraterrestrial environments \citep[e.g.][]{catlingdavid2018exoplanet, schwieterman2024overview,seager2012astrophysical}. 

{\bf Oxygen (O$_2$) and Ozone (O$_3$):} The most widely considered biosignature molecules for Earth-like planets are O$_2$ and its derivative O$_3$. O$_2$ is a prominent biogenic molecule in the Earth's atmosphere, with a volume mixing ratio of 21\%. Atmospheric O$_2$ on Earth is produced by oxygenic photosynthesis, in which a redox reaction between H$_2$O and CO$_2$ in the presence of sunlight results in O$_2$ and sugars. Photosynthesis occurs in both plants and marine microorganisms, e.g. such as green algae, with substantial O$_2$ contributions from both sources. On the other hand, O$_2$ is consumed by most complex life on Earth. O$_2$ has strong spectral signatures in the UV and visible wavelengths, with a strong band at 0.76 $\mu$m, along with weaker features in the infrared near 1.3 $\mu$m and 1.6 $\mu$m. O$_2$ satisfies all the requirements of an ideal biosignature for an Earth-like exoplanet due to its high abundance, strong spectral signatures, as well as a lack of significant abiotic sources on Earth. However, some studies have argued for the possibility of photochemical (abiotic) production of O$_2$ in a different planetary environment \citep{meadows2018exoplanet}.

O$_3$ is regarded as another good biosignature, as an indirect indicator of the presence of O$_2$, since on Earth it is produced directly from photochemical reactions involving O$_2$. The ozone layer plays an important role in the Earth's stratosphere by shielding the incoming solar UV radiation which can be harmful for life on Earth. O$_3$ can be depleted by reactions with various chemicals, a good example being chlorofluorocarbons. O$_3$ has strong features in the UV and visible range as well as in the infrared, near 3.3 $\mu$m, 4.7 $\mu$m and 9.6 $\mu$m, and is present in concentrations of a few parts per million (ppm) in the stratosphere. Similar to O$_2$, O$_3$ has also been considered as a good biosignature molecule for Earth-like planets but can also have photochemical abiotic origins. 

{\bf Methane (CH$_4$):} CH$_4$ has been a prominent biogenic gas in the atmosphere since very early in Earth's evolution, over 3.5 Gyr ago, well before the rise of O$_2$. It is produced on Earth primarily by microorganisms in a process called methanogenesis, whereby CO$_2$ is reduced to CH$_4$. The organisms producing CH$_4$, or methanogens, occur across the biosphere including both land and aquatic regions and in the digestive tracts of animals. It is important to note that methanogens can survive in anaerobic environments without the need for O$_2$ and are thought to be some of the first forms of simple microbial life on Earth before O$_2$ became abundant. In the present day atmosphere, CH$_4$ is present at $\sim$2 ppm abundance \citep{brasseur2005aeronomy} but is still one of the prominent greenhouse gases due to its strong infrared absorption. CH$_4$ has strong spectral features across the infrared, e.g. near at 1.4, 1.7, 2.3, 3.3 and 8 $\mu$m. CH$_4$ also has significant sinks in the atmosphere, e.g. direct photodissociation as well as destruction by reactions with oxygen radicals, such as the hydroxyl radical, resulting from photochemical destruction of molecules such as O$_3$ and H$_2$O. While CH$_4$ meets most of the requirements for an ideal biosignature for Earth-like planets, it is also produced in small quantities from geochemical reactions, e.g. serpentinization. Some CH$_4$ can also be produced through atmospheric chemistry for high enough concentrations of primordial H$_2$. Therefore, the interpretation of CH$_4$ as a biosignature relies both on its quantity and on other molecules detected in the atmosphere to provide the appropriate environmental context. 

{\bf Nitrous Oxide (N$_2$O):} Among the prominent biogenic gases in the Earth's atmosphere, N$_2$O is presently the most robust as it is not known to have any abiotic source. N$_2$O is produced predominantly by microorganisms as a byproduct of the nitrogen cycle, besides anthropogenic sources. While N is an important bioessential element, atmospheric N$_2$ is not readily usable by organisms. Specific microorganisms convert N$_2$ into usable compounds from which N is assimilated into other biota, and eventually released back into the atmosphere as N$_2$ and other byproducts, including N$_2$O. N$_2$O is also destroyed by photochemical reactions in the atmosphere. While a small amount of N$_2$O can also be produced abiotically, e.g. by lightning, it is not considered to be a significant source. The concentration of N$_2$O in the Earth's atmosphere from all sources is $\sim$0.3 ppm, but it is still one of the significant contributors to greenhouse warming due to its strong absorption in the infrared. N$_2$O has several prominent spectral features near 2.3, 2.9, 4.0 and 4.5 $\mu$m, and serves as a robust biosignature molecule for Earth-like planets. 

\subsubsection{Secondary Biosignature Gases} Besides the prominent molecules discussed above a number of other gases produced in small quantities also serve as unique biosignatures in Earth's atmosphere. Life on Earth produces hundreds of such molecules from various biotic processes \citep{seager2016toward}, but the vast majority of them are not abundant enough to be detectable in an exoplanetary atmosphere. At the same time, some of them may indeed be more abundant in exoplanetary atmospheres than on Earth. An exhaustive list of such molecules is beyond the scope of the present summary. However, a subset of such gases have been identified in the literature as particularly promising biosignatures to search for in exoplanets. We present some such examples here.

{\bf Sulphur-bearing gases:} A number of sulphur-bearing organic molecules have been identified as unique biosignatures in the Earth's atmosphere \citep{domagal2011using, schwieterman2024overview}. The most dominant sulphur-bearing biogenic molecule in the Earth's atmosphere is dimethyl sulphide (DMS), which is produced by microorganisms in terrestrial and aquatic environments, e.g. by marine phytoplankton and bacterial metabolism of some organic matter. DMS is particularly important in the Earth's atmosphere due to its role in the formation of sulphur-based aerosols in the atmosphere which act as nuclei for cloud formation and, in turn, affect the planetary albedo and climate. Other organosulphur molecules produced exclusively by life on Earth include dimethyl disulphide (DMDS) and methyl mercaptan or methanethiol (CH$_3$SH). Carbon disulphide (CS$_2$) is another potential biosignature but also has abiotic sources, e.g. volcanic outgassing. All these molecules are present in very small quantities and have short lifetimes in the Earth's atmosphere. For example, the prominent species DMS is present at an abundance below 1 part per billion (ppb) and a typical lifetime on the order of a day. These species are easily destroyed by free radicals, particularly OH, and photodissociation in the Earth's atmosphere. 

These molecules have been proposed as robust biosignatures on Earth-like planets as there are no known abiotic sources on Earth and they are difficult to retain in the atmosphere without a steady state biotic source. However, their low abundances make them more difficult to observe spectroscopically in Earth-like atmospheres. On the other hand, it has been suggested that these molecules could be detectable on larger exoplanets and those with H$_2$-rich atmospheres. All three molecules have strong spectral features in the infrared, e.g. DMS has strong features near 3.3 $\mu$m, 4.5 $\mu$m, 7 $\mu$m and 10 $\mu$m. Other sulphur-bearing molecules such as carbonyl sulphide (OCS) and carbon disulphide (CS$_2$) are also produced by life on Earth but they also have abiotic sources, such as volcanic outgassing. 

{\bf Methylated halogens:} Methylated halogens are known to be particularly good biosignatures because of their unique association with biotic processes \citep{leung2022ApJ}. They include methyl chloride (CH$_3$Cl) and methyl bromide (CH$_3$Br). Such gases are present in small quantities, below 1 ppb, but have strong spectroscopic features in the infrared. For example, CH$_3$Cl has strong features near 3.3 $\mu$m, 7 $\mu$m and 10 $\mu$m. Similar to other secondary biosignatures discussed above, the primary sinks for these organohalogens are also the OH radical and photodissociation in the atmosphere. These molecules have no significant abiotic sources on Earth, and are therefore considered to be strong biosignatures in exoplanetary environments where they could be present in larger abundances. Besides, CH$_3$Cl and CH$_3$Br, other organohalogens and methylated species in general can also act as good candidates for biosignatures in exoplanetary atmospheres \citep{schwieterman2024overview, leung2022ApJ}. 

{\bf Other hydrocarbons:} Besides CH$_4$, life on Earth produces many other hydrocarbons through numerous biochemical pathways. Among these, some key hydrocarbons have been identified as abundant enough and with strong enough spectral signatures to act as promising biosignatures from an astronomical perspective. These include molecules such as isoprene (C$_5$H$_8$) and ethane (C$_2$H$_6$). On Earth, isoprene is known to be a prominent molecule released by plants. On the other hand, ethane is a  strong byproduct of the conversion of DMS, and has, therefore, been proposed as a proxy for DMS \citep{domagal2011using}. 

{\bf Ammonia (NH$_3$):} On Earth, NH$_3$ is produced as a byproduct of the nitrogen cycle, e.g. by microorganisms which convert atmospheric N$_2$ into NH$_3$ or through bacterial disintegration of organic matter. Biogenic NH$_3$ in the atmosphere is present in smaller quantities compared to N$_2$O , with average mixing ratios up to few tens of ppb, and has a short lifetime due to its high solubility in water and its susceptibility to photodissociation. Therefore, NH$_3$ is not necessarily an ideal biosignature for Earth-like planets, but it can be abundant in other planetary envrionments, e.g. rocky planets with optimal H$_2$-rich atmospheres where it can have longer lifetimes and other production pathways \citep{seager2013exoplanet}. NH$_3$ also has strong spectral features across the infrared, making it particularly detectable for H$_2$-rich atmospheres. On the other hand, NH$_3$ may occur naturally due to thermochemical equilibrium in thick H$_2$-rich atmospheres, as seen in solar system giant planets at low temperatures, and can also be depleted in the presence of large surface oceans \citep{Hu2021, Tsai2021}, e.g. in hycean worlds \citep{madhusudhan2021habitability}. 

{\bf Phosphine (PH$_3$):} On Earth, PH$_3$ is uniquely associated with microbial environments but the biochemical pathways to its origins are not well understood. PH$_3$ is present in small quantities in the Earth, at mixing ratios below 1 ppm. Similar to several other biosignatures discussed above, PH$_3$ is susceptible to loss through reaction with oxygen radicals and photodissociation. Therefore, it is a better biosignature in oxygen-poor environments, including H$_2$-rich or CO$_2$-rich atmospheres. Similar to NH$_3$, PH$_3$ is also expected to have longer lifetimes in H$_2$-rich environments and has strong spectral signatures in the infrared. However, PH$_3$ can also be produced abiotically in thick H$_2$-rich atmospheres as seen in the atmospheres of solar system giant planets. Therefore, the interpretation of both NH$_3$ and PH$_3$ as biosignatures relies on the environmental context and finding other biosignature molecules for increased confidence in the potential presence of a biosphere. 

\subsection{Key Factors for Biosignature Assessment}
The assessment of a potential biosignature in a planetary atmosphere is very strongly tied to the environmental conditions. For example, while CH$_4$ is produced predominantly by life on Earth, it can be naturally abundant from thermochemical equilibrium in thick H$_2$-rich atmospheres, besides being produced from geological sources. Therefore, accurate assessment of a potential biosignature should take into account the overall atmospheric and possible surface conditions on a planet. A number of metrics have been suggested for assessing biosignatures, including considerations of environmental factors, chemical disequilibria, chemical network topologies and probabilistic approaches \citep{catlingdavid2018exoplanet, schwieterman2024overview}. Here, we discuss two key factors that are of central importance. 

\subsubsection{Environmental Context} For Earth-like planets orbiting sun-like stars, the prominent biosignatures on Earth such as O$_2$, O$_3$, CH$_4$, and N$_2$O would still serve as good signs of life. Even if one of these molecules may have a potential abiotic source it is unlikely that all of them together can be generated abiotically in significant quantities. For rocky planets orbiting M dwarfs, or otherwise UV active environments, it is possible that O$_2$ and O$_3$ may be generated abiotically, but the combination of two or more of the above molecules would still serve as a good biosignature. While the secondary biosignatures on Earth, such as DMS and CH$_3$Cl would serve as robust biosignatures, their abundances are expected to be low and difficult to observe in the atmospheres of small planets. 

For planets with H$_2$-rich atmospheres, the above molecules would likely not be strong biosignatures. In such a reduced environment, O$_2$ is not expected to be abundant. CH$_4$ can in principle still be contributed by life but will be difficult to disentangle from that produced abiotically. On the other hand, the secondary biosignatures such as DMS and CH$_3$Cl, could be more observable and hence take the role of dominant biosignatures. This is the case for both super-Earths with H$_2$-rich atmospheres \citep{seager2013exoplanet} as well as hycean worlds \citep{madhusudhan2021habitability}. Another potential biosignature in such environments is NH$_3$, but it can be underabundant in the presence of significant surface ocean as discussed above. While both CH$_4$ and NH$_3$ can be produced by thermochemistry in deep (e.g. $\gtrsim$100 bar) H$_2$-rich atmospheres, neither of them would be expected to be significantly abundant for thin (e.g. $\lesssim$ 10 bar) H$_2$-rich atmospheres and hence could be potential biosignatures in the latter case. Overall, a secure biosignature detection in any environment would rely on using multiple biosignature molecules given the environmental context. 

\subsubsection{Chemical Disequilibria}
Besides individual molecules, strong biosignatures can also involve specific pairs of molecules and indicators of biogenic chemical disequilibrium in an atmosphere. A central aspect of an ideal biosignature is a chemical signature that would not otherwise be produced by abiotic processes. One avenue to find such a signature is to find deviations in the abundances of prominent molecules from what can be explained by equilibrium and/or non-equilibrium processes for the given atmosphere. In particular, specific pairs of molecules have been proposed as such biosignatures in terrestrial-like atmospheres \citep{krissansen-Totton2018,schwieterman2024overview}. The CH$_4$ - CO$_2$ pair is one such example for terrestrial like atmospheres \citep{krissansen-Totton2018}. Even though CO$_2$ or CH$_4$ can each be produced abiotically to some extent, e.g. through volcanic emissions, the simultaneous presence of significant quantities of both in the absence of CO has been predicted to be a strong biosignature for Earth-like planets \citep{krissansen-Totton2018}. CH$_4$ is susceptible to short lifetimes in the atmosphere, both due to photodissociation and oxidation, and therefore needs a strong biogenic source to be present in significant quantities, e.g. mixing ratios $\gtrsim$ 10$^{-3}$. Similarly, disequilibrium combinations have also been suggested for CH$_4$-O$_2$, N$_2$-O$_2$, and N$_2$O-O$_2$-CH$_4$ as potential indicators of life on Earth-like planets.

Similar counterparts can also exist for non-Earth-like environments, e.g. those with H$_2$-rich atmospheres. For example, as discussed above, while CH$_4$ would be naturally abundant for deep H$_2$-rich atmospheres, the presence of significant CH$_4$ and CO$_2$ in the absence of CO in shallow H$_2$-rich atmospheres could be a potential  indicator of a biogenic source \citep{madhusudhan2023carbon, wogan2024jwst, cooke2024considerations}. Similarly, the detections of significant abundances of secondary terrestrial biosignatures in a H$_2$-rich atmosphere could also be important indicators. For example, molecules such as DMS are expected to have short lifetimes in both oxidised and reduced environments, due to photodissociation and reactions with radical species \citep{domagal2011using, seager2013exoplanet}. Therefore, their existence in significant quantities in a planetary atmosphere in the presence of other major species, e.g. oxidised species such as H$_2$O and CO$_2$, in steady state would require a sustained biogenic source.

\subsection{Surface and Temporal Biosignatures} In the above discussion, we have considered gaseous biosignatures which could be detected through the spectroscopic signatures of planetary atmospheres. This is the most accessible form of a biosignature. However, other approaches have been considered to detect spectroscopic signatures of surfaces or through temporal variations of specific molecules. The most common surface biosignature that has been predicted for Earth-like planets is due to the selective absorption of incident starlight by chlorophyll pigments in plants on the planetary surface. For example, green plants have strong absorption in the red part of the visible spectrum between $\sim$0.65-0.70 $\mu$m, but have a high reflectance between 0.70-0.75 $\mu$m. This leads to a 'red edge' in the reflected light spectrum near $\sim$0.70 $\mu$m, with lower albedo below this wavelength compared to that above \citep{seager2005}. Similar spectroscopic signatures are caused by a variety of plants and photosynthetic microorganisms which may be used as promising biosignatures in reflectance spectra indicative of surface vegetation. However, potential degeneracies may exist due to absorption and reflectance properties of some minerals on the planetary surface, and the nature of vegetation on exoplanets may be different due to different stellar spectra, e.g. of M dwarfs, which need to be considered while interpreting the data \citep{seager2005,schwieterman2024overview}. 

Another avenue for detecting potential biosignatures is through temporal variations in the atmospheric abundances of prominent gases caused by the biosphere \citep{schwieterman2024overview}. On Earth, the prominent gases O$_2$, CH$_4$, and CO$_2$ show variations in atmospheric concentrations over the year due to variation in growth and decay in vegetation and aquatic microorganisms. Such abundance variations could in turn show modulations in the absorption features of those gases in observed spectra over time which can be indicative of a biosignature.

\subsection{Current Status and Future Prospects}

The advent of large telescopes in space and on ground are beginning to make it possible to search for biosignatures on exoplanets. Several exoplanets are currently accessible to biosignature searches using transit spectroscopy with the James Webb Space Telescope (JWST), and many more with upcoming facilities using other methods, as follows. 

Current searches for biosignatures are focused on atmospheres of transiting exoplanets that are potentially habitable. The primary approach for probing such atmospheres is transmission spectroscopy, observed when the planet transits in front of the host star. As discussed in section~\ref{sec:habitable_environments}, the possibility of habitable environments has been predicted for different types of exoplanets, including rocky planets, ocean worlds and hycean worlds. Theoretical studies show that biosignatures may be detectable in these atmospheres with JWST with different levels of difficulty, Earth-like planets being the most difficult and hycean planets being the most accessible. The amplitude of a spectral feature in a transmission spectrum is higher for a larger planet size, smaller stellar size, and larger atmospheric scale height. The scale height in turn is proportional to the temperature and inversely proportional to the gravity and mean molecular weight (MMW). Therefore, larger and warmer planets with lighter atmospheres are more observable, especially around bright and small stars, e.g. nearby M dwarfs, which provide higher signal-to-noise ratio. Here we discuss the available targets, potential for biosignature detection, and current observations for different planet types.

Traditionally, the focus in the search for life has been on Earth-like habitable planets. Currently there is no Earth-size exoplanet known orbiting a sun-like star, and the atmospheres of such planets would be beyond the reach of current observational facilities, e.g. JWST. However, as discussed in section~\ref{sec:habitable_environments}, several Earth-size planets are known orbiting in the habitable-zones of low-mass, M dwarf, stars and some of them are accessible for atmospheric observations using JWST. The potential biosignatures in such atmospheres would be the prominent terrestrial biosginatures such as O$_2$, O$_3$, CH$_4$, N$_2$O and/or biosignature pairs such as CH$_4$-CO$_2$. Theoretical studies show that O$_3$ may be detected if present in the atmospheres of some of the TRAPPIST-1 planets using 30-60 transits with JWST, amounting to $\sim$100-200 hours per planet \citep{barstow2016habitable, lustig2019detectability}. The CH$_4$-CO$_2$ pair is expected to be detectable in $\sim$10 transits \citep{lustig2019detectability}. JWST observations of TRAPPIST-1 planets have not led to atmospheric detections so far and have highlighted challenges due to the very active star and the possibility that the planets may not have significant atmospheres. While such planets with comparable observability are currently limited to the TRAPPIST-1 system and a few more, the prospects will improve as more rocky planets are discovered around nearby M dwarfs.

At the other end of biosignature detectability are hycean planets, whose large sizes and H$_2$-rich atmospheres make them more conducive for atmospheric observations. The potential biosignatures in such atmospheres, as discussed above, are expected to be the secondary biosignatures such as DMS, DMDS, and/or CH$_3$Cl. While CH$_4$ can be another prominent biosignature it may also be produced by abiotic mechanisms in a H$_2$-rich atmosphere. Theoretical studies show that several such biosignatures can be detected in hycean worlds with a few transits, e.g. tens of hours, with JWST \citep{madhusudhan2021habitability}. The first detections for carbon-bearing molecules, CH$_4$ and CO$_2$, in a habitable-zone exoplanet were reported for the candidate hycean world K2-18~b \citep{madhusudhan2023carbon}. While the CH$_4$-CO$_2$ pair is a possible biosignature in terrestrial planets it is not yet known if it can be a robust biosignature in hycean worlds. The observations also provide tentative evidence for DMS in K2-18~b, but more observations are required to robustly establish its presence. About 15 candidate hycean worlds orbiting nearby M dwarfs are known as of 2025. 

Biosignatures may also be detectable with JWST in other habitable exoplanets, including ocean worlds with terrestrial-like atmospheres \citep{leger2004} and super-Earths with H$_2$-rich atmospheres \citep[e.g.][]{seager2013b_h2}. Their biosignatures are expected to be similar to those discussed for the terrestrial-like exoplanets and hycean worlds, respectively. There are currently no observational constraints on biosignatures on such exoplanets but they would be promising targets with JWST with similar considerations for their observability as discussed above. 

Beyond JWST, biosignatures on exoplanets will also be observable with other large telescopes in the future. Starting in the 2030s, the ground-based extremely large telescope (ELT) will be able to observe nearby habitable exoplanets. For transiting exoplanets, atmospheric signatures could be observed using high-resolution cross-correlation spectroscopy during transit. For non-transiting planets, such detections may be possible for some nearby exoplanets in reflected light using high-resolution spectroscopy and direct imaging. In the 2040s, searches for biosignatures in Earth-like planets orbiting sun-like stars would be possible in reflected light with the Habitable Worlds Observatory (HWO), a dedicated space mission to search for life on Earth-analogs, which is currently being planned. 

\section{Summary}\label{sec:summary}

The search for life beyond the solar system is the ultimate goal in exoplanetary science. Exoplanet surveys are increasingly detecting temperate low-mass exoplanets orbiting nearby stars and large observational facilities are beginning to probe the atmospheres of such planets. These developments are making the search for habitable environments and atmospheric biosignatures beyond the solar system more promising than ever before. The identification of a habitable planet and a biosignature in its environment needs consideration of a wide range of factors. Traditionally, these factors were considered primarily for Earth-like rocky planets around different stellar types. In recent years, there is a growing realization of the possibility of habitable conditions in environments very different from Earth. There has also been a growing understanding of the plausibility of different molecules as biosignatures in diverse environments. In this work, we provided an introduction to planetary habitability, the various factors affecting habitability, and the different types of habitable environments possible in exoplanetary systems. Similarly, we provided a broad overview of prominent and secondary biosignatures on Earth, their plausibility as potential biosignatures in diverse exoplanetary environments, and several factors of importance in assessing a biosignature. Importantly, some of the prominent biosignatures on Earth, such as O$_2$ and CH$_4$, may have abiotic sources in exoplanetary environments while some of the secondary terrestrial biosignatures may act as prominent biosigantures in some exoplanets. We discussed the current state of the art and future prospects in the search for biosignatures on exoplanets. 

\begin{ack}[Acknowledgments]

We thank Lorenzo Pica Ciamarra, Gregory Cooke, Savvas Constantinou, Edouard Barrier, and Charles Mawusi for proofreading and comments on the manuscript, and Charles Mawusi for help with reference collection and formatting of the manuscript.  

\end{ack}

\seealso{Extensive reviews on exoplanetary habitability can be found in \citet{cockell2016habitability,cockell2024sustained, Kaltenegger2017, meadows2018exoplanet, schulze2018, kane_book2021, catlingbook2017}, and those on biosignatures can be found in \citet{seager2016toward, catlingdavid2018exoplanet, schwieterman2018exoplanet, schwieterman2024overview}.}

\bibliographystyle{Harvard}
\bibliography{reference}

\begin{thebibliography*}{48}
\providecommand{\bibtype}[1]{}
\providecommand{\natexlab}[1]{#1}
{\catcode`\|=0\catcode`\#=12\catcode`\@=11\catcode`\\=12
|immediate|write|@auxout{\expandafter\ifx\csname
  natexlab\endcsname\relax\gdef\natexlab#1{#1}\fi}}
\renewcommand{\url}[1]{{\tt #1}}
\providecommand{\urlprefix}{URL }
\expandafter\ifx\csname urlstyle\endcsname\relax
  \providecommand{\doi}[1]{doi:\discretionary{}{}{}#1}\else
  \providecommand{\doi}{doi:\discretionary{}{}{}\begingroup
  \urlstyle{rm}\Url}\fi
\providecommand{\bibinfo}[2]{#2}
\providecommand{\eprint}[2][]{\url{#2}}

\bibtype{Article}%
\bibitem[{Anglada-Escud{\'e}} et al.(2016)]{anglada2016}
\bibinfo{author}{{Anglada-Escud{\'e}} G}, \bibinfo{author}{{Amado} PJ},
  \bibinfo{author}{{Barnes} J}, \bibinfo{author}{{Berdi{\~n}as} ZM},
  \bibinfo{author}{{Butler} RP}, \bibinfo{author}{{Coleman} GAL},
  \bibinfo{author}{{de La Cueva} I}, \bibinfo{author}{{Dreizler} S},
  \bibinfo{author}{{Endl} M}, \bibinfo{author}{{Giesers} B},
  \bibinfo{author}{{Jeffers} SV}, \bibinfo{author}{{Jenkins} JS},
  \bibinfo{author}{{Jones} HRA}, \bibinfo{author}{{Kiraga} M},
  \bibinfo{author}{{K{\"u}rster} M}, \bibinfo{author}{{L{\'o}pez-Gonz{\'a}lez}
  MJ}, \bibinfo{author}{{Marvin} CJ}, \bibinfo{author}{{Morales} N},
  \bibinfo{author}{{Morin} J}, \bibinfo{author}{{Nelson} RP},
  \bibinfo{author}{{Ortiz} JL}, \bibinfo{author}{{Ofir} A},
  \bibinfo{author}{{Paardekooper} SJ}, \bibinfo{author}{{Reiners} A},
  \bibinfo{author}{{Rodr{\'\i}guez} E},
  \bibinfo{author}{{Rodr{\'\i}guez-L{\'o}pez} C}, \bibinfo{author}{{Sarmiento}
  LF}, \bibinfo{author}{{Strachan} JP}, \bibinfo{author}{{Tsapras} Y},
  \bibinfo{author}{{Tuomi} M} and  \bibinfo{author}{{Zechmeister} M}
  (\bibinfo{year}{2016}), \bibinfo{month}{Aug.}
\bibinfo{title}{{A terrestrial planet candidate in a temperate orbit around
  Proxima Centauri}}.
\bibinfo{journal}{{\em \nat}} \bibinfo{volume}{536} (\bibinfo{number}{7617}):
  \bibinfo{pages}{437--440}. \bibinfo{doi}{\doi{10.1038/nature19106}}.
\eprint{1609.03449}.

\bibtype{Article}%
\bibitem[Barstow and Irwin(2016)]{barstow2016habitable}
\bibinfo{author}{Barstow JK} and  \bibinfo{author}{Irwin PG}
  (\bibinfo{year}{2016}).
\bibinfo{title}{Habitable worlds with jwst: transit spectroscopy of the
  trappist-1 system?}
\bibinfo{journal}{{\em Monthly Notices of the Royal Astronomical Society:
  Letters}} \bibinfo{volume}{461} (\bibinfo{number}{1}):
  \bibinfo{pages}{L92--L96}.

\bibtype{Book}%
\bibitem[Brasseur and Solomon(2005)]{brasseur2005aeronomy}
\bibinfo{author}{Brasseur GP} and  \bibinfo{author}{Solomon S}
  (\bibinfo{year}{2005}).
\bibinfo{title}{Aeronomy of the middle atmosphere: Chemistry and physics of the
  stratosphere and mesosphere}, \bibinfo{volume}{32},
  \bibinfo{publisher}{Springer Science \& Business Media}.

\bibtype{Book}%
\bibitem[{Catling} and {Kasting}(2017)]{catlingbook2017}
\bibinfo{author}{{Catling} DC} and  \bibinfo{author}{{Kasting} JF}
  (\bibinfo{year}{2017}).
\bibinfo{title}{{Atmospheric Evolution on Inhabited and Lifeless Worlds}}.

\bibtype{Article}%
\bibitem[Catling et al.(2018)]{catlingdavid2018exoplanet}
\bibinfo{author}{Catling DC}, \bibinfo{author}{Kiang NY},
  \bibinfo{author}{Robinson TD}, \bibinfo{author}{Rushby AJ},
  \bibinfo{author}{Del~Genio A} and  et al. (\bibinfo{year}{2018}).
\bibinfo{title}{Exoplanet biosignatures: a framework for their assessment}.
\bibinfo{journal}{{\em Astrobiology}} .

\bibtype{Article}%
\bibitem[Cockell et al.(2016)]{cockell2016habitability}
\bibinfo{author}{Cockell CS}, \bibinfo{author}{Bush T}, \bibinfo{author}{Bryce
  C}, \bibinfo{author}{Direito S}, \bibinfo{author}{Fox-Powell M},
  \bibinfo{author}{Harrison JP}, \bibinfo{author}{Lammer H},
  \bibinfo{author}{Landenmark H}, \bibinfo{author}{Martin-Torres J},
  \bibinfo{author}{Nicholson N} and  et al. (\bibinfo{year}{2016}).
\bibinfo{title}{Habitability: a review}.
\bibinfo{journal}{{\em Astrobiology}} \bibinfo{volume}{16}
  (\bibinfo{number}{1}): \bibinfo{pages}{89--117}.

\bibtype{Article}%
\bibitem[Cockell et al.(2024)]{cockell2024sustained}
\bibinfo{author}{Cockell CS}, \bibinfo{author}{Simons M},
  \bibinfo{author}{Castillo-Rogez J}, \bibinfo{author}{Higgins PM},
  \bibinfo{author}{Kaltenegger L}, \bibinfo{author}{Keane JT},
  \bibinfo{author}{Leonard EJ}, \bibinfo{author}{Mitchell KL},
  \bibinfo{author}{Park RS}, \bibinfo{author}{Perl SM} and  et al.
  (\bibinfo{year}{2024}).
\bibinfo{title}{Sustained and comparative habitability beyond earth}.
\bibinfo{journal}{{\em Nature Astronomy}} \bibinfo{volume}{8}
  (\bibinfo{number}{1}): \bibinfo{pages}{30--38}.

\bibtype{Article}%
\bibitem[Cooke and Madhusudhan(2024)]{cooke2024considerations}
\bibinfo{author}{Cooke GJ} and  \bibinfo{author}{Madhusudhan N}
  (\bibinfo{year}{2024}).
\bibinfo{title}{Considerations for photochemical modeling of possible hycean
  worlds}.
\bibinfo{journal}{{\em arXiv preprint arXiv:2410.07313}} .

\bibtype{Article}%
\bibitem[Domagal-Goldman et al.(2011)]{domagal2011using}
\bibinfo{author}{Domagal-Goldman SD}, \bibinfo{author}{Meadows VS},
  \bibinfo{author}{Claire MW} and  \bibinfo{author}{Kasting JF}
  (\bibinfo{year}{2011}).
\bibinfo{title}{Using biogenic sulfur gases as remotely detectable
  biosignatures on anoxic planets}.
\bibinfo{journal}{{\em Astrobiology}} \bibinfo{volume}{11}
  (\bibinfo{number}{5}): \bibinfo{pages}{419--441}.

\bibtype{inbook}%
\bibitem[Egli(2009)]{egli2009}
\bibinfo{author}{Egli T} (\bibinfo{year}{2009}), \bibinfo{month}{01}.
\bibinfo{title}{Nutrition, Microbial}.
\bibinfo{comment}{ISBN} \bibinfo{isbn}{9780123739445},
  \bibinfo{pages}{308--324}.
\bibinfo{doi}{\doi{10.1016/B978-012373944-5.00083-3}}.

\bibtype{Article}%
\bibitem[{Gandhi} et al.(2020)]{gandhi2020}
\bibinfo{author}{{Gandhi} S}, \bibinfo{author}{{Brogi} M},
  \bibinfo{author}{{Yurchenko} SN}, \bibinfo{author}{{Tennyson} J},
  \bibinfo{author}{{Coles} PA}, \bibinfo{author}{{Webb} RK},
  \bibinfo{author}{{Birkby} JL}, \bibinfo{author}{{Guilluy} G},
  \bibinfo{author}{{Hawker} GA}, \bibinfo{author}{{Madhusudhan} N},
  \bibinfo{author}{{Bonomo} AS} and  \bibinfo{author}{{Sozzetti} A}
  (\bibinfo{year}{2020}), \bibinfo{month}{Jun.}
\bibinfo{title}{{Molecular cross-sections for high-resolution spectroscopy of
  super-Earths, warm Neptunes, and hot Jupiters}}.
\bibinfo{journal}{{\em \mnras}} \bibinfo{volume}{495} (\bibinfo{number}{1}):
  \bibinfo{pages}{224--237}. \bibinfo{doi}{\doi{10.1093/mnras/staa981}}.
\eprint{2004.04160}.

\bibtype{Article}%
\bibitem[Gillon et al.(2017)]{gillon2017seven}
\bibinfo{author}{Gillon M}, \bibinfo{author}{Triaud AH},
  \bibinfo{author}{Demory BO}, \bibinfo{author}{Jehin E}, \bibinfo{author}{Agol
  E}, \bibinfo{author}{Deck KM}, \bibinfo{author}{Lederer SM},
  \bibinfo{author}{De~Wit J}, \bibinfo{author}{Burdanov A},
  \bibinfo{author}{Ingalls JG} and  et al. (\bibinfo{year}{2017}).
\bibinfo{title}{Seven temperate terrestrial planets around the nearby ultracool
  dwarf star trappist-1}.
\bibinfo{journal}{{\em Nature}} \bibinfo{volume}{542} (\bibinfo{number}{7642}):
  \bibinfo{pages}{456--460}.

\bibtype{Article}%
\bibitem[Gordon et al.(2017)]{HITRAN2016}
\bibinfo{author}{Gordon I}, \bibinfo{author}{Rothman L}, \bibinfo{author}{Hill
  C}, \bibinfo{author}{Kochanov R}, \bibinfo{author}{Tan Y},
  \bibinfo{author}{Bernath P}, \bibinfo{author}{Birk M},
  \bibinfo{author}{Boudon V}, \bibinfo{author}{Campargue A},
  \bibinfo{author}{Chance K}, \bibinfo{author}{Drouin B},
  \bibinfo{author}{Flaud JM}, \bibinfo{author}{Gamache R},
  \bibinfo{author}{Hodges J}, \bibinfo{author}{Jacquemart D},
  \bibinfo{author}{Perevalov V}, \bibinfo{author}{Perrin A},
  \bibinfo{author}{Shine K}, \bibinfo{author}{Smith MA},
  \bibinfo{author}{Tennyson J}, \bibinfo{author}{Toon G},
  \bibinfo{author}{H.~Tran VT}, \bibinfo{author}{Barbe A},
  \bibinfo{author}{Csaszar A}, \bibinfo{author}{Devi V},
  \bibinfo{author}{Furtenbacher T}, \bibinfo{author}{Harrison J},
  \bibinfo{author}{Hartmann JM}, \bibinfo{author}{Jolly A},
  \bibinfo{author}{Johnson T}, \bibinfo{author}{Karman T},
  \bibinfo{author}{Kleiner I}, \bibinfo{author}{Kyuberis A},
  \bibinfo{author}{Loos J}, \bibinfo{author}{Lyulin O}, \bibinfo{author}{Massie
  S}, \bibinfo{author}{Mikhailenko S}, \bibinfo{author}{Moazzen-Ahmadi N},
  \bibinfo{author}{Muller H}, \bibinfo{author}{Naumenko O},
  \bibinfo{author}{Nikitin A}, \bibinfo{author}{Polyansky O},
  \bibinfo{author}{Rey M}, \bibinfo{author}{Rotger M}, \bibinfo{author}{Sharpe
  S}, \bibinfo{author}{Sung K}, \bibinfo{author}{Starikova E},
  \bibinfo{author}{Tashkun S}, \bibinfo{author}{Auwera JV},
  \bibinfo{author}{Wagner G}, \bibinfo{author}{Wilzewski J},
  \bibinfo{author}{Wcislo P}, \bibinfo{author}{Yu S} and  \bibinfo{author}{Zak
  E} (\bibinfo{year}{2017}).
\bibinfo{title}{The hitran2016 molecular spectroscopic database}.
\bibinfo{journal}{{\em Journal of Quantitative Spectroscopy and Radiative
  Transfer}} \bibinfo{doi}{\doi{10.1016/j.jqsrt.2017.06.038}}.

\bibtype{Article}%
\bibitem[{Gronoff} et al.(2020)]{gronoff2020}
\bibinfo{author}{{Gronoff} G}, \bibinfo{author}{{Arras} P},
  \bibinfo{author}{{Baraka} S}, \bibinfo{author}{{Bell} JM},
  \bibinfo{author}{{Cessateur} G}, \bibinfo{author}{{Cohen} O},
  \bibinfo{author}{{Curry} SM}, \bibinfo{author}{{Drake} JJ},
  \bibinfo{author}{{Elrod} M}, \bibinfo{author}{{Erwin} J},
  \bibinfo{author}{{Garcia-Sage} K}, \bibinfo{author}{{Garraffo} C},
  \bibinfo{author}{{Glocer} A}, \bibinfo{author}{{Heavens} NG},
  \bibinfo{author}{{Lovato} K}, \bibinfo{author}{{Maggiolo} R},
  \bibinfo{author}{{Parkinson} CD}, \bibinfo{author}{{Simon Wedlund} C},
  \bibinfo{author}{{Weimer} DR} and  \bibinfo{author}{{Moore} WB}
  (\bibinfo{year}{2020}), \bibinfo{month}{Aug.}
\bibinfo{title}{{Atmospheric Escape Processes and Planetary Atmospheric
  Evolution}}.
\bibinfo{journal}{{\em Journal of Geophysical Research (Space Physics)}}
  \bibinfo{volume}{125} (\bibinfo{number}{8}), \bibinfo{eid}{e27639}.
  \bibinfo{doi}{\doi{10.1029/2019JA02763910.1002/essoar.10502458.1}}.

\bibtype{Article}%
\bibitem[{Hu} et al.(2021)]{Hu2021}
\bibinfo{author}{{Hu} R}, \bibinfo{author}{{Damiano} M},
  \bibinfo{author}{{Scheucher} M}, \bibinfo{author}{{Kite} E},
  \bibinfo{author}{{Seager} S} and  \bibinfo{author}{{Rauer} H}
  (\bibinfo{year}{2021}), \bibinfo{month}{Nov.}
\bibinfo{title}{{Unveiling Shrouded Oceans on Temperate sub-Neptunes via
  Transit Signatures of Solubility Equilibria versus Gas Thermochemistry}}.
\bibinfo{journal}{{\em \apjl}} \bibinfo{volume}{921} (\bibinfo{number}{1}),
  \bibinfo{eid}{L8}.

\bibtype{Article}%
\bibitem[{Kaltenegger}(2017)]{Kaltenegger2017}
\bibinfo{author}{{Kaltenegger} L} (\bibinfo{year}{2017}), \bibinfo{month}{Aug.}
\bibinfo{title}{{How to Characterize Habitable Worlds and Signs of Life}}.
\bibinfo{journal}{{\em \araa}} \bibinfo{volume}{55} (\bibinfo{number}{1}):
  \bibinfo{pages}{433--485}.
  \bibinfo{doi}{\doi{10.1146/annurev-astro-082214-122238}}.
\eprint{1911.05597}.

\bibtype{Book}%
\bibitem[{Kane}(2021)]{kane_book2021}
\bibinfo{author}{{Kane} SR} (\bibinfo{year}{2021}).
\bibinfo{title}{{Planetary Habitability}}.
\bibinfo{doi}{\doi{10.1088/2514-3433/ac2aa1}}.

\bibtype{Article}%
\bibitem[Kasting et al.(1993)]{kasting1993habitable}
\bibinfo{author}{Kasting JF}, \bibinfo{author}{Whitmire DP} and
  \bibinfo{author}{Reynolds RT} (\bibinfo{year}{1993}).
\bibinfo{title}{Habitable zones around main sequence stars}.
\bibinfo{journal}{{\em Icarus}} \bibinfo{volume}{101} (\bibinfo{number}{1}):
  \bibinfo{pages}{108--128}.

\bibtype{Article}%
\bibitem[{Kopparapu} et al.(2013)]{Kopparapu2013}
\bibinfo{author}{{Kopparapu} RK}, \bibinfo{author}{{Ramirez} R},
  \bibinfo{author}{{Kasting} JF}, \bibinfo{author}{{Eymet} V},
  \bibinfo{author}{{Robinson} TD}, \bibinfo{author}{{Mahadevan} S},
  \bibinfo{author}{{Terrien} RC}, \bibinfo{author}{{Domagal-Goldman} S},
  \bibinfo{author}{{Meadows} V} and  \bibinfo{author}{{Deshpande} R}
  (\bibinfo{year}{2013}), \bibinfo{month}{Mar.}
\bibinfo{title}{{Habitable Zones around Main-sequence Stars: New Estimates}}.
\bibinfo{journal}{{\em \apj}} \bibinfo{volume}{765} (\bibinfo{number}{2}),
  \bibinfo{eid}{131}. \bibinfo{doi}{\doi{10.1088/0004-637X/765/2/131}}.
\eprint{1301.6674}.

\bibtype{Article}%
\bibitem[{Krissansen-Totton} et al.(2018)]{krissansen-Totton2018}
\bibinfo{author}{{Krissansen-Totton} J}, \bibinfo{author}{{Olson} S} and
  \bibinfo{author}{{Catling} DC} (\bibinfo{year}{2018}), \bibinfo{month}{Jan.}
\bibinfo{title}{{Disequilibrium biosignatures over Earth history and
  implications for detecting exoplanet life}}.
\bibinfo{journal}{{\em Science Advances}} \bibinfo{volume}{4}
  (\bibinfo{number}{1}): \bibinfo{pages}{eaao5747}.
  \bibinfo{doi}{\doi{10.1126/sciadv.aao5747}}.
\eprint{1801.08211}.

\bibtype{incollection}%
\bibitem[Lammer(2012)]{lammer2012escape}
\bibinfo{author}{Lammer H} (\bibinfo{year}{2012}), \bibinfo{title}{Escape of
  planetary atmospheres}, \bibinfo{booktitle}{Origin and Evolution of Planetary
  Atmospheres: Implications for Habitability}, \bibinfo{publisher}{Springer},
  \bibinfo{pages}{25--74}.

\bibtype{Article}%
\bibitem[{Lauro} et al.(2021)]{lauro2021}
\bibinfo{author}{{Lauro} SE}, \bibinfo{author}{{Pettinelli} E},
  \bibinfo{author}{{Caprarelli} G}, \bibinfo{author}{{Guallini} L},
  \bibinfo{author}{{Rossi} AP}, \bibinfo{author}{{Mattei} E},
  \bibinfo{author}{{Cosciotti} B}, \bibinfo{author}{{Cicchetti} A},
  \bibinfo{author}{{Soldovieri} F}, \bibinfo{author}{{Cartacci} M},
  \bibinfo{author}{{Di Paolo} F}, \bibinfo{author}{{Noschese} R} and
  \bibinfo{author}{{Orosei} R} (\bibinfo{year}{2021}), \bibinfo{month}{Jan.}
\bibinfo{title}{{Multiple subglacial water bodies below the south pole of Mars
  unveiled by new MARSIS data}}.
\bibinfo{journal}{{\em Nature Astronomy}} \bibinfo{volume}{5}:
  \bibinfo{pages}{63--70}. \bibinfo{doi}{\doi{10.1038/s41550-020-1200-6}}.
\eprint{2010.00870}.

\bibtype{Article}%
\bibitem[{L{\'e}ger} et al.(2004)]{leger2004}
\bibinfo{author}{{L{\'e}ger} A}, \bibinfo{author}{{Selsis} F},
  \bibinfo{author}{{Sotin} C}, \bibinfo{author}{{Guillot} T},
  \bibinfo{author}{{Despois} D}, \bibinfo{author}{{Mawet} D},
  \bibinfo{author}{{Ollivier} M}, \bibinfo{author}{{Lab{\`e}que} A},
  \bibinfo{author}{{Valette} C}, \bibinfo{author}{{Brachet} F},
  \bibinfo{author}{{Chazelas} B} and  \bibinfo{author}{{Lammer} H}
  (\bibinfo{year}{2004}), \bibinfo{month}{Jun.}
\bibinfo{title}{{A new family of planets? ``Ocean-Planets''}}.
\bibinfo{journal}{{\em \icarus}} \bibinfo{volume}{169} (\bibinfo{number}{2}):
  \bibinfo{pages}{499--504}. \bibinfo{doi}{\doi{10.1016/j.icarus.2004.01.001}}.
\eprint{astro-ph/0308324}.

\bibtype{Article}%
\bibitem[{Leung} et al.(2022)]{leung2022ApJ}
\bibinfo{author}{{Leung} M}, \bibinfo{author}{{Schwieterman} EW},
  \bibinfo{author}{{Parenteau} MN} and  \bibinfo{author}{{Fauchez} TJ}
  (\bibinfo{year}{2022}), \bibinfo{month}{Oct.}
\bibinfo{title}{{Alternative Methylated Biosignatures. I. Methyl Bromide, a
  Capstone Biosignature}}.
\bibinfo{journal}{{\em \apj}} \bibinfo{volume}{938} (\bibinfo{number}{1}),
  \bibinfo{eid}{6}. \bibinfo{doi}{\doi{10.3847/1538-4357/ac8799}}.
\eprint{2208.07393}.

\bibtype{Article}%
\bibitem[Lustig-Yaeger et al.(2019)]{lustig2019detectability}
\bibinfo{author}{Lustig-Yaeger J}, \bibinfo{author}{Meadows VS} and
  \bibinfo{author}{Lincowski AP} (\bibinfo{year}{2019}).
\bibinfo{title}{The detectability and characterization of the trappist-1
  exoplanet atmospheres with jwst}.
\bibinfo{journal}{{\em The Astronomical Journal}} \bibinfo{volume}{158}
  (\bibinfo{number}{1}): \bibinfo{pages}{27}.

\bibtype{Article}%
\bibitem[{Madhusudhan}(2019)]{Madhusudhan2019}
\bibinfo{author}{{Madhusudhan} N} (\bibinfo{year}{2019}), \bibinfo{month}{Aug.}
\bibinfo{title}{{Exoplanetary Atmospheres: Key Insights, Challenges, and
  Prospects}}.
\bibinfo{journal}{{\em \araa}} \bibinfo{volume}{57}: \bibinfo{pages}{617--663}.
  \bibinfo{doi}{\doi{10.1146/annurev-astro-081817-051846}}.
\eprint{1904.03190}.

\bibtype{Article}%
\bibitem[Madhusudhan et al.(2021)]{madhusudhan2021habitability}
\bibinfo{author}{Madhusudhan N}, \bibinfo{author}{Piette AA} and
  \bibinfo{author}{Constantinou S} (\bibinfo{year}{2021}).
\bibinfo{title}{Habitability and biosignatures of hycean worlds}.
\bibinfo{journal}{{\em The Astrophysical Journal}} \bibinfo{volume}{918}
  (\bibinfo{number}{1}): \bibinfo{pages}{1}.

\bibtype{Article}%
\bibitem[Madhusudhan et al.(2023{\natexlab{a}})]{madhusudhan2023chemical}
\bibinfo{author}{Madhusudhan N}, \bibinfo{author}{Moses JI},
  \bibinfo{author}{Rigby F} and  \bibinfo{author}{Barrier E}
  (\bibinfo{year}{2023}{\natexlab{a}}).
\bibinfo{title}{Chemical conditions on hycean worlds}.
\bibinfo{journal}{{\em Faraday Discussions}} \bibinfo{volume}{245}:
  \bibinfo{pages}{80--111}.

\bibtype{Article}%
\bibitem[Madhusudhan et al.(2023{\natexlab{b}})]{madhusudhan2023carbon}
\bibinfo{author}{Madhusudhan N}, \bibinfo{author}{Sarkar S},
  \bibinfo{author}{Constantinou S}, \bibinfo{author}{Holmberg M},
  \bibinfo{author}{Piette AA} and  \bibinfo{author}{Moses JI}
  (\bibinfo{year}{2023}{\natexlab{b}}).
\bibinfo{title}{Carbon-bearing molecules in a possible hycean atmosphere}.
\bibinfo{journal}{{\em The Astrophysical Journal Letters}}
  \bibinfo{volume}{956} (\bibinfo{number}{1}): \bibinfo{pages}{L13}.

\bibtype{Article}%
\bibitem[Meadows and Barnes(2018)]{meadows2018factors}
\bibinfo{author}{Meadows VS} and  \bibinfo{author}{Barnes RK}
  (\bibinfo{year}{2018}).
\bibinfo{title}{Factors affecting exoplanet habitability}.
\bibinfo{journal}{{\em Handbook of exoplanets}} : \bibinfo{pages}{57}.

\bibtype{Article}%
\bibitem[Meadows et al.(2018)]{meadows2018exoplanet}
\bibinfo{author}{Meadows VS}, \bibinfo{author}{Reinhard CT},
  \bibinfo{author}{Arney GN}, \bibinfo{author}{Parenteau MN},
  \bibinfo{author}{Schwieterman EW}, \bibinfo{author}{Domagal-Goldman SD},
  \bibinfo{author}{Lincowski AP}, \bibinfo{author}{Stapelfeldt KR},
  \bibinfo{author}{Rauer H}, \bibinfo{author}{DasSarma S} and  et al.
  (\bibinfo{year}{2018}).
\bibinfo{title}{Exoplanet biosignatures: understanding oxygen as a biosignature
  in the context of its environment}.
\bibinfo{journal}{{\em Astrobiology}} \bibinfo{volume}{18}
  (\bibinfo{number}{6}): \bibinfo{pages}{630--662}.

\bibtype{Article}%
\bibitem[{Merino} et al.(2019)]{merino2019}
\bibinfo{author}{{Merino} N}, \bibinfo{author}{{Aronson} HS},
  \bibinfo{author}{{Bojanova} DP}, \bibinfo{author}{{Feyhl-Buska} J},
  \bibinfo{author}{{Wong} ML}, \bibinfo{author}{{Zhang} S} and
  \bibinfo{author}{{Giovannelli} D} (\bibinfo{year}{2019}),
  \bibinfo{month}{Apr.}
\bibinfo{title}{{Living at the Extremes: Extremophiles and the Limits of Life
  in a Planetary Context}}.
\bibinfo{journal}{{\em Frontiers in Microbiology}} \bibinfo{volume}{10}:
  \bibinfo{pages}{780}.
  \bibinfo{doi}{\doi{10.3389/fmicb.2019.0078010.31223/osf.io/8eay6}}.

\bibtype{Article}%
\bibitem[{Pierrehumbert} and {Gaidos}(2011)]{Pierrehumbert2011}
\bibinfo{author}{{Pierrehumbert} R} and  \bibinfo{author}{{Gaidos} E}
  (\bibinfo{year}{2011}), \bibinfo{month}{Jun.}
\bibinfo{title}{{Hydrogen Greenhouse Planets Beyond the Habitable Zone}}.
\bibinfo{journal}{{\em \apjl}} \bibinfo{volume}{734} (\bibinfo{number}{1}),
  \bibinfo{eid}{L13}.
  \bibinfo{doi}{\doi{10.1088/2041-8205/734/1/L1310.48550/arXiv.1105.0021}}.
\eprint{1105.0021}.

\bibtype{Article}%
\bibitem[{Rothschild} and {Mancinelli}(2001)]{Rothschild2001}
\bibinfo{author}{{Rothschild} LJ} and  \bibinfo{author}{{Mancinelli} RL}
  (\bibinfo{year}{2001}), \bibinfo{month}{Feb.}
\bibinfo{title}{{Life in extreme environments}}.
\bibinfo{journal}{{\em \nat}} \bibinfo{volume}{409} (\bibinfo{number}{6823}):
  \bibinfo{pages}{1092--1101}. \bibinfo{doi}{\doi{10.1038/35059215}}.

\bibtype{Article}%
\bibitem[{Rushby} et al.(2013)]{rushby2013}
\bibinfo{author}{{Rushby} AJ}, \bibinfo{author}{{Claire} MW},
  \bibinfo{author}{{Osborn} H} and  \bibinfo{author}{{Watson} AJ}
  (\bibinfo{year}{2013}), \bibinfo{month}{Sep.}
\bibinfo{title}{{Habitable Zone Lifetimes of Exoplanets around Main Sequence
  Stars}}.
\bibinfo{journal}{{\em Astrobiology}} \bibinfo{volume}{13}
  (\bibinfo{number}{9}): \bibinfo{pages}{833--849}.
  \bibinfo{doi}{\doi{10.1089/ast.2012.0938}}.

\bibtype{Book}%
\bibitem[{Schulze-Makuch} and {Irwin}(2018)]{schulze2018}
\bibinfo{author}{{Schulze-Makuch} D} and  \bibinfo{author}{{Irwin} LN}
  (\bibinfo{year}{2018}).
\bibinfo{title}{{Life in the Universe}}, \bibinfo{publisher}{Springer Cham}.
\bibinfo{doi}{\doi{10.1007/978-3-319-97658-7}}.

\bibtype{Article}%
\bibitem[Schwieterman and Leung(2024)]{schwieterman2024overview}
\bibinfo{author}{Schwieterman EW} and  \bibinfo{author}{Leung M}
  (\bibinfo{year}{2024}).
\bibinfo{title}{An overview of exoplanet biosignatures}.
\bibinfo{journal}{{\em arXiv preprint arXiv:2404.15431}} .

\bibtype{Article}%
\bibitem[Schwieterman et al.(2018)]{schwieterman2018exoplanet}
\bibinfo{author}{Schwieterman EW}, \bibinfo{author}{Kiang NY},
  \bibinfo{author}{Parenteau MN}, \bibinfo{author}{Harman CE},
  \bibinfo{author}{DasSarma S}, \bibinfo{author}{Fisher TM},
  \bibinfo{author}{Arney GN}, \bibinfo{author}{Hartnett HE},
  \bibinfo{author}{Reinhard CT}, \bibinfo{author}{Olson SL} and  et al.
  (\bibinfo{year}{2018}).
\bibinfo{title}{Exoplanet biosignatures: a review of remotely detectable signs
  of life}.
\bibinfo{journal}{{\em Astrobiology}} \bibinfo{volume}{18}
  (\bibinfo{number}{6}): \bibinfo{pages}{663--708}.

\bibtype{Article}%
\bibitem[Seager(2013)]{seager2013exoplanet}
\bibinfo{author}{Seager S} (\bibinfo{year}{2013}).
\bibinfo{title}{Exoplanet habitability}.
\bibinfo{journal}{{\em Science}} \bibinfo{volume}{340}
  (\bibinfo{number}{6132}): \bibinfo{pages}{577--581}.

\bibtype{Article}%
\bibitem[{Seager} et al.(2005)]{seager2005}
\bibinfo{author}{{Seager} S}, \bibinfo{author}{{Turner} EL},
  \bibinfo{author}{{Schafer} J} and  \bibinfo{author}{{Ford} EB}
  (\bibinfo{year}{2005}), \bibinfo{month}{Jun.}
\bibinfo{title}{{Vegetation's Red Edge: A Possible Spectroscopic Biosignature
  of Extraterrestrial Plants}}.
\bibinfo{journal}{{\em Astrobiology}} \bibinfo{volume}{5}
  (\bibinfo{number}{3}): \bibinfo{pages}{372--390}.
  \bibinfo{doi}{\doi{10.1089/ast.2005.5.372}}.
\eprint{astro-ph/0503302}.

\bibtype{Article}%
\bibitem[Seager et al.(2012)]{seager2012astrophysical}
\bibinfo{author}{Seager S}, \bibinfo{author}{Schrenk M} and
  \bibinfo{author}{Bains W} (\bibinfo{year}{2012}).
\bibinfo{title}{An astrophysical view of earth-based metabolic biosignature
  gases}.
\bibinfo{journal}{{\em Astrobiology}} \bibinfo{volume}{12}
  (\bibinfo{number}{1}): \bibinfo{pages}{61--82}.

\bibtype{Article}%
\bibitem[{Seager} et al.(2013)]{seager2013b_h2}
\bibinfo{author}{{Seager} S}, \bibinfo{author}{{Bains} W} and
  \bibinfo{author}{{Hu} R} (\bibinfo{year}{2013}), \bibinfo{month}{Nov.}
\bibinfo{title}{{Biosignature Gases in H$_{2}$-dominated Atmospheres on Rocky
  Exoplanets}}.
\bibinfo{journal}{{\em \apj}} \bibinfo{volume}{777} (\bibinfo{number}{2}),
  \bibinfo{eid}{95}. \bibinfo{doi}{\doi{10.1088/0004-637X/777/2/95}}.
\eprint{1309.6016}.

\bibtype{Article}%
\bibitem[Seager et al.(2016)]{seager2016toward}
\bibinfo{author}{Seager S}, \bibinfo{author}{Bains W} and
  \bibinfo{author}{Petkowski J} (\bibinfo{year}{2016}).
\bibinfo{title}{Toward a list of molecules as potential biosignature gases for
  the search for life on exoplanets and applications to terrestrial
  biochemistry}.
\bibinfo{journal}{{\em Astrobiology}} \bibinfo{volume}{16}
  (\bibinfo{number}{6}): \bibinfo{pages}{465--485}.

\bibtype{Article}%
\bibitem[Selsis et al.(2007)]{selsis2007habitable}
\bibinfo{author}{Selsis F}, \bibinfo{author}{Kasting JF},
  \bibinfo{author}{Levrard B}, \bibinfo{author}{Paillet J},
  \bibinfo{author}{Ribas I} and  \bibinfo{author}{Delfosse X}
  (\bibinfo{year}{2007}).
\bibinfo{title}{Habitable planets around the star gliese 581?}
\bibinfo{journal}{{\em Astronomy \& Astrophysics}} \bibinfo{volume}{476}
  (\bibinfo{number}{3}): \bibinfo{pages}{1373--1387}.

\bibtype{Article}%
\bibitem[{Tennyson} et al.(2016)]{ExoMol2016}
\bibinfo{author}{{Tennyson} J}, \bibinfo{author}{{Yurchenko} SN},
  \bibinfo{author}{{Al-Refaie} AF}, \bibinfo{author}{{Barton} EJ},
  \bibinfo{author}{{Chubb} KL}, \bibinfo{author}{{Coles} PA},
  \bibinfo{author}{{Diamantopoulou} S}, \bibinfo{author}{{Gorman} MN},
  \bibinfo{author}{{Hill} C}, \bibinfo{author}{{Lam} AZ},
  \bibinfo{author}{{Lodi} L}, \bibinfo{author}{{McKemmish} LK},
  \bibinfo{author}{{Na} Y}, \bibinfo{author}{{Owens} A},
  \bibinfo{author}{{Polyansky} OL}, \bibinfo{author}{{Rivlin} T},
  \bibinfo{author}{{Sousa-Silva} C}, \bibinfo{author}{{Underwood} DS},
  \bibinfo{author}{{Yachmenev} A} and  \bibinfo{author}{{Zak} E}
  (\bibinfo{year}{2016}), \bibinfo{month}{Sep.}
\bibinfo{title}{{The ExoMol database: Molecular line lists for exoplanet and
  other hot atmospheres}}.
\bibinfo{journal}{{\em Journal of Molecular Spectroscopy}}
  \bibinfo{volume}{327}: \bibinfo{pages}{73--94}.
  \bibinfo{doi}{\doi{10.1016/j.jms.2016.05.002}}.
\eprint{1603.05890}.

\bibtype{Article}%
\bibitem[{Tsai} et al.(2021)]{Tsai2021}
\bibinfo{author}{{Tsai} SM}, \bibinfo{author}{{Innes} H},
  \bibinfo{author}{{Lichtenberg} T}, \bibinfo{author}{{Taylor} J},
  \bibinfo{author}{{Malik} M}, \bibinfo{author}{{Chubb} K} and
  \bibinfo{author}{{Pierrehumbert} R} (\bibinfo{year}{2021}),
  \bibinfo{month}{Dec.}
\bibinfo{title}{{Inferring Shallow Surfaces on Sub-Neptune Exoplanets with
  JWST}}.
\bibinfo{journal}{{\em \apjl}} \bibinfo{volume}{922} (\bibinfo{number}{2}),
  \bibinfo{eid}{L27}.

\bibtype{Article}%
\bibitem[Wogan et al.(2024)]{wogan2024jwst}
\bibinfo{author}{Wogan NF}, \bibinfo{author}{Batalha NE},
  \bibinfo{author}{Zahnle KJ}, \bibinfo{author}{Krissansen-Totton J},
  \bibinfo{author}{Tsai SM} and  \bibinfo{author}{Hu R} (\bibinfo{year}{2024}).
\bibinfo{title}{Jwst observations of k2-18b can be explained by a gas-rich
  mini-neptune with no habitable surface}.
\bibinfo{journal}{{\em The Astrophysical Journal Letters}}
  \bibinfo{volume}{963} (\bibinfo{number}{1}): \bibinfo{pages}{L7}.

\bibtype{Article}%
\bibitem[Yang et al.(2013)]{yang2013stabilizing}
\bibinfo{author}{Yang J}, \bibinfo{author}{Cowan NB} and
  \bibinfo{author}{Abbot DS} (\bibinfo{year}{2013}).
\bibinfo{title}{Stabilizing cloud feedback dramatically expands the habitable
  zone of tidally locked planets}.
\bibinfo{journal}{{\em The Astrophysical Journal Letters}}
  \bibinfo{volume}{771} (\bibinfo{number}{2}): \bibinfo{pages}{L45}.

\end{thebibliography*}

\end{document}